\documentclass[
,showpacs ,amssymb,nobibnotes, aps,eqsecnum]{revtex4}
\usepackage{graphicx}
\input{epsf}

\usepackage{amsmath,amssymb}
\usepackage{bm}

\newcommand{\dalm}{\kern1pt\vbox{\hrule height 0.9pt\hbox{\vrule width 0.9pt
\hskip 2.5pt\vbox{\vskip 5.5pt}\hskip 3pt\vrule width 0.3pt}\hrule height 0.3pt}
\kern1pt}

\begin{document}



\title{Gravitational Radiation from Collapsing Magnetized Dust. II \\
-- Polar Parity Perturbation --}

\author{Hajime Sotani}\email{sotani@astro.auth.gr}
\affiliation{
%
%
%
Theoretical Astrophysics, University of T\"{u}bingen, Auf der Morgenstelle 10,
T\"{u}bingen 72076, Germany
}

\date{\today}

\begin{abstract}
With the gauge-invariant perturbation theory, we study the effects of the stellar magnetic fields
on the polar gravitational waves emitted during the homogeneous dust collapse.
We found that the emitted energy in gravitational waves depends strongly not only on the initial stellar
radius but also on the rasio between the poloidal and toroidal magnetic components.
The polar gravitational wave output of such a collapse can be easily up to a few order of magnitude larger
than what we get from the nonmagnetized collapse. The changes due to the existence of magnetic field could
be helpful to extract some information of inner magnetic profiles of progenitor from the detection of 
the gravitational waves radiated during the black hole formation, which results from the stellar collapse.
\end{abstract}

\maketitle
\section{Introduction}
\label{sec:I}

The direct observation of gravitational waves is a significant attempt
in theoretical and experimental physics. In order to accomplish this great purpose,
several ground-based laser interferometric detectors with kilometer-size arm,
such as LIGO, TAMA300, GEO600, and VIRGO, are currently operating and the next-generation
detectors are also on the menu \cite{Barish2005}. In addition to the ground-based detectors,
the projects to launch a detector in space, such as LISA \cite{LISA} and
DECIGO \cite{DECIGO}, are in progress.
The importance to detect gravitational waves directly is that one can obtain a new method
to see the Universe by gravitational waves, which is called ``gravitational-wave astronomy".
In fact, by using the gravitational waves associated with the oscillations of compact object,
it is possible to determine the stellar radius, mass, equation of state (EOS), and so on
\cite{Andersson1996,Kokkotas2001,Sotani2002,Sotani2003,Sotani2004,Sotani2009a,Erich2008,Wolfgang2008}.
Additionally, with collecting the observational data of gravitational waves,
we may able to verify the gravitational theory, and new physics at a high density or
high energy region.

For both ground-based and space detectors, the nonspherical stellar collapse is one of the most
promising sources of gravitational waves. With their sensitivity, the black hole formation with
stellar mass is a target for the ground-based detectors, while the space detectors might detect
signals radiated from the creation of supermassive black holes \cite{sp1985,ss2002,Baiotti2005}.
The most hopeful approach to calculate these gravitational waves emitted during the black hole
formation, is the numerical relativity, i.e., via direct numerical integration of the exact
Einstein and hydrodynamic equations, which are full-nonlinearly coupled with each other.
In the last decade, the numerical relativity made dramatic developments and it becomes possible
to treat many complicated matter and spacetime \cite{Duez2006,Baiotti2006,Dimmelmeier2007,Anderson2008}.
Still, the computation of gravitational waves with high accuracy is not technically easy, because
the gravitational waves emitted from the stellar collapse are very weak and sometimes they could contain
unphysical noises due to gauge modes and/or numerical error. So in this paper, as an alternative
approach, we consider the linear perturbation theory. This is possible to extract the weak
gravitational waves with precision and also becomes a cross-check for the numerical results
with numerical relativity.

With respect to the calculation of gravitational waves radiated from the stellar collapse to the black
hole, initially, Cunningham, Price and Moncrief derived the perturbation equations on the Oppenheimer-Snyder
solution, which describes a homogeneous dust collapse \cite{OS1939}, and calculated radiated gravitational waves
\cite{Cunningham1978}. Subsequently, Seidel and co-workers studied the gravitational waves emitted from
the stellar collapse when a neutron star would be born \cite{Seidel1987}, where they used the
gauge-invariant perturbation formalism on the spherically symmetric spacetime formulated by Gerlach and
Sengupta \cite{Gerlach1979}. Iguchi, Nakao and Harada investigated nonspherical perturbations of a collapsing
inhomogeneous dust ball \cite{Iguchi1998}, which is described by the Lema\^{\i}tre-Tolman-Bondi solution
\cite{ltb1933}. Further, Harada, Iguchi and Shibata calculated the axial gravitational waves emitted from
the collapse of a supermassive star to a black hole by employing the covariant gauge-invariant formalism
on the spherically symmetric spacetime and the coordinate-independent matching conditions at stellar surface,
which is devised by Gundlach and Mart\'{\i}n-Garc\'{\i}a \cite{Gundlach2000}. Recently, with the same
formalism, Sotani, Yoshida and Kokkotas considered the magnetic effect on the axial gravitational waves
emitted from the collapse of homogeneous dust sphere \cite{SYK2007}
(hereafter, we refer to this article as Paper I).

In spite of many investigations of gravitational radiation from the stellar collapse with linear
perturbation analysis as mentioned the above, they have not included the effect of the magnetic
fields on the emitted gravitational waves, except for Paper I. It should notice that Cunningham,
Price and Moncrief also dealt with the electromagnetic perturbations on the Oppenheimer-Snyder solution,
but they did not consider the direct coupling between fluid and magnetic field \cite{Cunningham1978}.
While, the importance of magnetic effects on the evolution of compact objects has been recently realized
due to the appearance of new instruments with high performance. One of the most remarkable examples
is the discovery of magnetars, which are neutron stars with strong magnetic field such as $B > 10^{15}$
Gauss. With magnetar models, it is successful to explain the observed specific frequencies of
quasi-periodic oscillations (QPOs) in the decaying tail of the giant flares \cite{Sotani2006,Sotani2008,Sotani2009b}.
Since there exists a quite strong magnetic field in some neutron stars, which would be produced after
the stellar collapse, it is natural to take into account its effect on stellar collapse. And it could
be also probable that the magnetic fields of the collapsing object are amplified during the collapse
due to the magnetic flux conservation and they would affect the emitted gravitational waves, even
if the initial magnetic field is weak. Actually, Paper I shows the possibility that the magnetic fields
affect on the axial gravitational waves emitted during the dust collapse. Additionally, there exists
another example showing the importance of magnetic effects in the evolution of compact objects, which is
related to gamma ray bursts (GRBs), i.e., the short-duration GRBs could arise from hypergiant flares
of magnetars associated with the soft gamma repeaters (SGRs) \cite{Nakar2006} or collapse of
magnetized hypermassive neutron star \cite{Shibata2006}.

Indeed, all examples mentioned above suggest that the magnetic fields play an important role in the stellar
collapse and its effect should not be negligible. So in this paper, in order to explore the effects of magnetic
fields on the emitted gravitational waves for the black hole formation, we consider the polar gravitational
waves radiated during the collapse of homogeneous dust ball with weak magnetic field. In particular,
along with Paper I, we focus on the only quadrupole gravitation waves, which could be more important
in the astrophysical point of view. The weak magnetic fields are treated as small perturbations on the
Oppenheimer-Snyder solution and we make an investigation with the covariant gauge-invariant formalism
on the spherically symmetric spacetime and the coordinate-independent matching conditions at stellar surface
proposed by Gundlach and Mart\'{\i}n-Garc\'{\i}a \cite{Gundlach2000}. It should be emphasized that so far there
is no calculation of polar gravitational waves on the dynamical background spacetime with the same method,
i.e., this paper is first calculation. The reason why the polar gravitational waves could not be solved,
might be the difficulty to deal with the boundary conditions at the stellar surface due to the existence
of many perturbative variables, in contrast to the axial gravitational waves. Furthermore, as shown
in the main text for ordering of perturbations, we consider the first order perturbations for the metric
and fluid motion while the second order perturbations for magnetic fields, to see the magnetic effects
on the emitted gravitational waves.

This paper is organized as follows. In Sec. \ref{sec:II} we briefly describe the gauge-invariant
perturbation theory on the spherical symmetric background, the background solution that we
adopt in this paper, and how to introduce the magnetic fields as the perturbations.
Next, in Sec. \ref{sec:III}, we derive the perturbation equations for polar gravitational waves
emitted during the collapse of magnetized dust sphere. Then the details of numerical procedure are
shown in Sec. \ref{sec:IV}, and we devote Sec. \ref{sec:V} to description of the code tests.
In Sec. \ref{sec:VI}, we show the numerical results related to the influence of existence of magnetic fields
on the gravitational waves radiated during the formation of black hole.
Finally we make a conclusion in Sec. \ref{sec:VII}.
In this paper, we adopt the unit of $c=G=1$, where $c$
and $G$ denote the speed of light and the gravitational constant, respectively, and
the metric signature is $(-,+,+,+)$.

\section{Basic Properties}
\label{sec:II}

As with Paper I, we deal with the electromagnetic fields as the small perturbations
on the dust sphere, since the magnetic energy is much smaller that the gravitational
binding energy even if the source of gravitational waves would involve a strong magnetic field
like a magnetar.
Thus the background metric $g_{\mu\nu}$ and four-velocity of fluid $u^{\mu}$ are determined
as solutions of a collapsing spherical dust sphere without the electromagnetic fields.
Now it is convenient to introduce two small dimensionless
parameters related to strength of the magnetic field and to amplitude of the 
gravitational waves, i.e., $\eta\sim|B/(GM^2 R_s^{-4})^{1/2}|$ and 
$\epsilon\sim |\delta g_{\mu\nu}|$, where $R_s$ is stellar radius and
we assume that the fluid perturbations
are also small, i.e., $|\delta u^{\mu}|\sim \epsilon$ and $|\delta \rho | \sim \epsilon$.
Then the leading terms for the perturbations of $t_{\mu\nu}^{(M)}$
and $t_{\mu\nu}^{(EM)}$ are $\delta t_{\mu\nu}^{(E)}\sim {\cal O}(\epsilon)$
and $\delta t_{\mu\nu}^{(EM)}\sim {\cal O}(\eta^2)$, where $t_{\mu\nu}^{(M)}$ and $t_{\mu\nu}^{(EM)}$
express the energy-momentum tensors for the fluid and for the electromagnetic fields, respectively.
In this paper since we focus on the effect of magnetic field upon the emitted gravitational waves during
the stellar collapse, we omit terms of higher order such as ${\cal O}(\epsilon^2)$ and ${\cal O}(\epsilon^1\eta^2)$.
Further with assumption that $\epsilon\sim\eta^2$, 
the perturbed Einstein equations of order $\epsilon$ are reduced to the following form
\begin{equation}
\delta G_{\mu\nu} = 8\pi\{\delta t^{(M)}_{\mu\nu}+\delta t^{(EM)}_{\mu\nu}\}
                    +{\cal O}(\epsilon^2)
                  = 8\pi\delta t_{\mu\nu} +{\cal O}(\epsilon^2) \,.
\end{equation}
Notice in this approximation the gravitational perturbations are driven both by the magnetic field 
and the fluid motions of the collapsing dust sphere.

\subsection{Gauge-Invariant Perturbation Theory}
\label{sec:II-1}

For spherically symmetric background spacetime,
the first order gauge-invariant perturbation theory has been formulated
by Gerlach and Sengupta \cite{Gerlach1979} and further developed
by Gundlach and Mart\'{\i}n-Garc\'{\i}a \cite{Gundlach2000}. 
In this subsection we briefly describe this formalism for the polar parity perturbations.

\subsubsection{Background Spacetime}
\label{sec:II-1-1}

The background spacetime, which is
a spherically symmetric four dimensional spacetime ${\cal M}$, can be described
as a product of the form  ${\cal M}={\cal M}^2 \times {\cal S}^2$, 
where ${\cal M}^2$ is a 2-dimensional (1+1) reduced spacetime and 
${\cal S}^2$ a 2-dimensional spheres. 
In other words, the metric $g_{\mu\nu}$ and the stress-energy tensor $t_{\mu\nu}$
on ${\cal M}$ can be written in the form
\begin{align}
 g_{\mu\nu} &\equiv \mbox{diag} (g_{AB},r^2\gamma_{ab}), \\
 t_{\mu\nu} &\equiv \mbox{diag} (t_{AB},Qr^2\gamma_{ab}),
\end{align}
where $g_{AB}$ is an arbitrary ($1+1$) Lorentzian metric on
${\cal M}^2$, $r$ a scalar on ${\cal M}^2$, $Q$ some function on ${\cal M}^2$ 
and $\gamma_{ab}$  is the unit curvature metric on ${\cal S}^2$.
Note that  if the background spacetime is spherically symmetric then $Q=t^a_{\ a}/2$ .
Here and henceforth the Greek indices denote the spacetime components, 
the capital Latin indices the ${\cal M}^2$ components, and the small Latin indices
are used to denote the ${\cal S}^2$ components. 
Furthermore, the covariant derivatives on ${\cal M}$, ${\cal M}^2$, and
${\cal S}^2$ are represented by $_{;\mu}$, $_{|A}$, and $_{:a}$, respectively.
Finally, the totally antisymmetric covariant unit tensor on ${\cal M}^2$ 
is denoted as $\varepsilon_{AB}$ and on ${\cal S}^2$  as $\varepsilon_{ab}$.

\subsubsection{Nonradial Perturbations}
\label{sec:II-1-2}

As mentioned before, in this paper, we consider axisymmetric polar parity
perturbations both for the metric $\delta g_{\mu\nu}$ and the matter
perturbations $\delta t_{\mu\nu}$, which are given by
\begin{align}
 \delta g_{\mu\nu} & \equiv   \left(
   \begin{array}{cc}
       h_{AB}Y^{lm} &  h_A^{\rm (p)} Y^{lm}_{\ \ :a} \\
       *         &  r^2 (KY^{lm} \gamma_{ab} + GY^{lm}_{\ \ :ab})
   \end{array}\right), \label{PMP} \\
 \delta t_{\mu\nu} & \equiv   \left(
   \begin{array}{cc}
       \Delta t_{AB}Y^{lm} &  \Delta t_A^{\rm (p)} Y^{lm}_{\ \ :a} \\
       *                &  r^2 \Delta t^3Y^{lm} \gamma_{ab} + \Delta t^2 Z_{ab}^{lm}
   \end{array}\right)\,, \label{PFP}
\end{align}
where $Z_{ab}^{lm}\equiv Y^{lm}_{\ \ :ab} + l(l+1)Y^{lm} \gamma_{ab}/2$ and $Y^{lm}$ stands for the 
spherical harmonics.
With $h_{AB}$, $h_A^{\rm (p)}$, $K$, $G$, $\Delta t_{AB}$, $\Delta t_{A}^{\rm (p)}$, $\Delta t^3$, and $\Delta t^2$,
the gauge-invariant variables for the nonradial perturbations are defined as
\begin{align}
 k_{AB} &\equiv h_{AB}-(p_{A|B} +p_{B|A}), \\
 k      &\equiv K - 2v^A p_A, \\
 T_{AB} &\equiv \Delta t_{AB} - t_{AB|C}p^C - t_{AC}p_{\ |B}^C - t_{BC}p_{\ |A}^C, \\
 T_A    &\equiv \Delta t_{A}^{\rm (p)} - t_{AC}p^C - \frac{r^2}{2}QG_{|A}, \\
 T^2    &\equiv \Delta t^2 - r^2QG, \\
 T^3    &\equiv \Delta t^3 - (Q_{|C} + 2Qv_C)p^C + \frac{l(l+1)}{2}QG,
\end{align}
where $v_A\equiv r_{|A}/r$ and $p_A\equiv h_A^{\rm (p)} - r^2 G_{|A}/2$ \cite{Gundlach2000}.
Note that $T_A$ is defined for $l\ge 1$ and $T^2$ for $l\ge 2$.
In terms of the gauge-invariant variables, the linearized Einstein equations for the polar 
parity perturbations are given by \cite{Gerlach1979};
\begin{align}
  &2v^C\left(k_{AB|C} - k_{CA|B} - k_{CB|A}\right)
      - \left[\frac{l(l+1)}{r^2} + G_C^{\ C} + G_a^{\ a} + 2{\cal R}\right]k_{AB}
      - 2g_{AB}v^C \left(k_{ED|C} - k_{CE|D} - k_{CD|E}\right)g^{ED} \nonumber \\
  &\hspace{2cm}+ g_{AB}\left(2v^{C|D} + 4v^C v^D - G^{CD}\right)k_{CD}
      + g_{AB} \left[\frac{l(l+1)}{r^2}
      + \frac{1}{2}\left(G_C^{\ C} + G_a^{\ a}\right) + {\cal R}\right]k_D^{\ \ D}  \nonumber \\
  &\hspace{2cm}+ 2\left(v_A k_{|B} + v_B k_{|A} + k_{|A|B}\right)
      - g_{AB} \left[2k_{|C}^{\ \ |C} + 6v^C k_{|C} - \frac{(l-1)(l+2)}{r^2}k\right]
      = -16 \pi T_{AB}, \label{GS1} \\
  &k_{|A} - k_{AB}^{\ \ \ |B} + k_{B\ \ |A}^{\ \ B} - v_A k_B^{\ \ B} = -16 \pi T_A, \label{GS2} \\
  &\left(k_{|A}^{\ \ |A} + 2v^A k_{|A} + G_a^{\ a}k\right)
      - \left[k_{AB}^{\ \ \ |A|B} + 2v^A k_{AB}^{\ \ \ |B} + 2\left(v^{A|B} + v^A v^B\right)k_{AB}\right] \nonumber \\
  &\hspace{2cm}+ \left[k_{A\ \ |B}^{\ \ A\ \ |B} + v^A k_{B\ \ |A}^{\ \ B} + {\cal R}k_A^{\ \ A}
      - \frac{l(l+1)}{2r^2}k_A^{\ \ A}\right] = 16 \pi T^3, \label{GS3} \\
  &k_A^{\ A} = -16\pi T^2, \label{GS4}
\end{align}
where ${\cal R}$ is the Gaussian curvature on ${\cal M}^2$, and $G_{AB}$ and $G_a^{\ a}$ are defined as
\begin{align}
  G_{AB} &\equiv -2\left(v_{A|B} + v_Av_B\right) + g_{AB} V_0, \\
  G_a^{\ a} &\equiv 2\left(v_A^{\ |A} + v_A v^A - {\cal R}\right),
\end{align}
where $V_0$ is defined as
$V_0\equiv 2\left(-\dot{U} + W' - \mu U + \nu W\right) + 3\left(W^2 - U^2\right) - r^{-2}$
and $\mu\equiv u^A_{\ \ |A}$, $\nu\equiv n^A_{\ \ |A}$, $U\equiv u^A v_A$, $W\equiv n^A v_A$,
$\dot{F}\equiv u^A F_{|A}$, and $F' \equiv n^AF_{|A}$ \cite{MG1999}.
Now if the symmetric tensor $k_{AB}$ is decomposed with a coordinate-independent way into three scalars,
such as
\begin{equation}
  k_{AB} \equiv q(-u_A u_B + n_A n_B) + \phi (u_A u_B + n_A n_B) + \psi (u_A n_B + n_A u_B),
\end{equation}
and to eliminate $\phi$ we introduce the new variable $\zeta$ defined as $\zeta \equiv \phi - k + q$,
then from Eqs. (\ref{GS1}) -- (\ref{GS4})
we can get the perturbation equations for the variables of metric perturbations as
\begin{gather}
  -\ddot{\zeta} + \zeta'' + 2(\mu - U)\psi' = S_\zeta, \label{perturbation-01} \\
  -\ddot{k} + c_s^2 k'' - 2c_s^2 U\psi' = S_k, \label{perturbation-02} \\
  -\dot{\psi} = S_\psi, \label{perturbation-03} \\
  q = -8\pi T^2, \label{perturbation-04}
\end{gather}
where the source terms $S_\zeta$, $S_k$, and $S_\psi$ are given in Appendix \ref{sec:appendix_1},
which are without the variables for matter perturbations.
Namely, we can calculate the evolutions for metric perturbations independent of the matter perturbations.
On the other hand, the variables associated with the matter perturbations are given by
\begin{gather}
  -8\pi n^A u^B T_{AB} = (\dot{k})' + C_{\gamma}, \label{perturbation-05} \\
  8\pi u^A u^B T_{AB} = - k'' + 2 U\psi' + C_{\omega}, \label{perturbation-06} \\
  -16\pi u^A T_A = \psi' + C_{\alpha}. \label{perturbation-07}
\end{gather}
Notice that the right hand sides of Eqs. (\ref{perturbation-05}) -- (\ref{perturbation-07})
are produced with only variables for metric perturbations, i.e., the matter perturbations
can be determined after the calculation for the metric perturbations.
The concrete forms of $C_\gamma$, $C_\omega$, and $C_\alpha$ are described in Appendix \ref{sec:appendix_1}.

\subsection{Oppenheimer-Snyder Solution}
\label{sec:II-2}

We briefly describe the adopted background spacetime which will be later 
endowed with a  magnetic field.
We consider perturbations around a homogeneous spherically symmetric dust 
collapse described by the Oppenheimer-Snyder (OS) solution, whose line 
element inside the dust sphere is given by
\begin{align}
ds^2 
  &=-d\tau^2+R^2(\tau)[d\chi^2 + \sin^2 \chi (d \theta^2 +\sin^2\theta d \phi^2)]\,,
\label{OSM} \\
  &= R^2(\eta)[-d\eta^2+d\chi^2+ \sin^2 \chi (d \theta^2 +\sin^2\theta d \phi^2)]\,,
\label{OSM1}
\end{align}
where $\chi$ is a radial coordinate defined in the range of 
$0\le\chi\le\chi_0<\pi/2$ and $\chi_0$ corresponds to the stellar surface.
Additionally $R(\eta)$ and $\tau(\eta)$ are the scale factor and the proper time of a comoving observer with the fluid,
respectively, which are defined in terms of the conformal time $\eta$ as follows
\begin{align}
 R(\eta)    &= \frac{M}{\sin^3 \chi_0}(1 + \cos\eta)\,, \\
 \tau(\eta) &= \frac{M}{\sin^3 \chi_0}(\eta + \sin\eta)\,,
\end{align}
where $M$ is the total gravitational mass of the dust sphere.
The energy-momentum tensor for the dust fluid is written as
\begin{equation}
 t_{\mu\nu}^{(M)} = \rho u_{\mu}u_{\nu},
\end{equation}
where $\rho$ is the rest mass density given by
\begin{equation}
 \rho (\eta) = \frac{3 \sin^6\chi_0}{4\pi M^2}(1+\cos\eta)^{-3}\,,
\end{equation}
and $u^{\mu}$ denotes the four-velocity of the dust, described in  terms of
comoving coordinates as
%
\begin{equation}
 u^\mu=\delta^\mu_{\ \tau}\, \ \ \ \mbox{or}\ \ \ u^\mu=R(\eta)^{-1}\delta^\mu_{\ \eta}
\end{equation}
where $\delta^\mu_{\ \nu}$ means the Kronecker delta.
Also with the four-velocity $u^\mu$, the spacelike unit vector defined as $n^{A}\equiv -\varepsilon_{AB}u^{B}$
is given by
\begin{equation}
 n^A=R(\eta)^{-1}\delta^{A}_{\ \chi},
\end{equation}
where $n^A$ is a normal vector on the sphere whose radius is constant.
Thus we have $\mu=U=\partial_{\eta}R/R^2$, $\nu=0$, and $W=\cos\chi/(R\sin\chi)$.
Furthermore the frame derivatives are $\dot{F}=\partial_\eta F/R$ and $F'=\partial_\chi F/R$.
The spacetime outside the dust sphere is described by the Schwarzschild metric, 
i.e.,
\begin{equation}
 ds^2 = -f(r)dt^2 + f(r)^{-1}dr^2 + r^2(d \theta^2 + \sin^2\theta d \phi^2), \label{Sch}
\end{equation}
where $f(r)\equiv 1-2M/r$.
Note that with the Schwarzschild metric
$\tilde{u}^A=(1/\sqrt{f},0)$, $\tilde{n}^A=(0,\sqrt{f})$, $\tilde{U}=\tilde{\mu}=0$,
$\tilde{W}=\sqrt{f}/r$, and $\tilde{\nu}=M/(r^2\sqrt{f})$,
where we use the tilde to avoid mixing of the variables inside the star.
From the junction conditions at the surface of the dust sphere, we obtain
the relationships between the $(\eta,\chi)$-coordinates and the $(t,r)$-coordinates, given by
\begin{align}
 r_s &= R(\eta)\sin\chi_0, \\
 \frac{t}{2M} &= \ln \left|\frac{[(r_{s0}/2M)-1]^{1/2}+\tan(\eta/2)}{[(r_{s0}/2M)-1]^{1/2}-\tan(\eta/2)}\right|
    + \left(\frac{r_{s0}}{2M}-1\right)^{1/2}\left[\eta + \left(\frac{r_{s0}}{4M}\right)(\eta + \sin\eta)\right],
\end{align}
where $r_{s0}\equiv r_s(t=0) = 2M/\sin^2\chi_0$ is the initial stellar radius 
in Schwarzschild coordinates.

\subsection{Magnetic Fields}
\label{sec:II-3}

As mentioned the above, we consider weakly magnetized dust spheres in which the 
magnetic effects are treated as small perturbations on the OS solution,
where we can consider that the electromagnetic fields are axisymmetric due to the nature of
a spherical symmetric background. Thus, perturbations of the
electromagnetic fields, $\delta F_{\mu\nu}$, can be described in terms of the spherical harmonics
$Y^{lm}$ by the following relations
\begin{align}
 \delta F_{01} &=  - \delta F_{10} =  e_2 Y^{lm}\,, \label{F-1} \\
 \delta F_{0a} &=  - \delta F_{a0} = e_1 S^{lm}_{\ \ a} + e_3 Y^{lm}_{\ \ :a} \,, \label{F-2} \\
 \delta F_{1a} &=  - \delta F_{a1} = b_1 S^{lm}_{\ \ a} + b_3 Y^{lm}_{\ \ :a} \,, \label{F-3} \\
 \delta F_{23} &=  - \delta F_{32} = b_2 \varepsilon_{23} Y^{lm}\,, \label{F-4}
%
\end{align}
where $S^{lm}_{\ \ a}\equiv \varepsilon^b_{\ a}Y^{lm}_{\ \ :b}$.
These variables of $\delta F_{\mu\nu}$ are governed by the Maxwell equations, i.e.,
\begin{gather}
\delta F_{\mu\nu,\sigma} +\delta F_{\nu\sigma,\mu} +\delta F_{\sigma\mu,\nu} = 0, 
\label{Maxwell-1} \\
\delta F^{\mu\nu}_{\ \ ;\nu} = 4\pi \delta J^{\mu}, \label{Maxwell-2}
\end{gather}
where $\delta J^{\mu}$ is the perturbations of the current four-vector. 
The equations (\ref{Maxwell-1}) and (\ref{Maxwell-2}) are correct up to 
order of $\epsilon^0\eta^1$.

In the interior of the dust sphere, we consider infinitely conductive fluids,
i.e., the ideal magnetohydrodynamic approximation has been adopted, 
according to which $\delta F_{\mu\nu}u^{\nu}=0$.
With this approximation and Maxwell equation (\ref{Maxwell-1}),
we obtain the basic equations for electromagnetic fields inside the star such as
\begin{gather}
 e_1 = e_2 = e_3 =0\,. \label{Maxwell1-0in} \\
 \partial_{\eta} b_1 = \partial_{\eta} b_2 = \partial_{\eta} b_3 = 0, \label{Maxwell1-1in} \\
 l(l+1)b_1 + \partial_{\chi} {b_2} = 0\,. \label{Maxwell1-2in}
\end{gather}
The equation (\ref{Maxwell1-1in}) tells us that the magnetic field does not change for a comoving observer.
Additionally, with the Faraday tensor $F^{\mu\nu}$, since the magnetic field, $B_\mu$, can be described as
$B_\mu = \epsilon_{\mu\nu\alpha\beta}u^{\nu}F^{\alpha\beta}/2$,
one can see that the quantities $b_1$ and $b_2$ are related
to the poloidal magnetic component while $b_3$ is associated with the toroidal magnetic component.
Furthermore, as similar to Paper I, in this paper we consider the case
that the magnetic fields are confined in the stellar interior, i.e., $b_1=b_2=b_3=0$ at stellar surface.
(In fact the necessary condition that the magnetic fields are confined inside the star is only $b_2=0$
at stellar surface, which is derived from the junction conditions for the magnetic fields as Paper I,
but for simplicity in this paper we adopt the above conditions.)
Of course the electromagnetic fields outside the star are also important to see the emission of
electro-magnetic waves, but they can be seen in the near future elsewhere.
Additionally in this paper we focus only on dipole electromagnetic fields,
i.e., electromagnetic fields associated with $l= 1$.
In the next section we will see how the dipole electromagnetic fields can drive the quadrupole gravitational 
radiation, which is more important in the observation.

\section{Perturbation Equations for polar parity}
\label{sec:III}

\subsection{Interior region of the star}
\label{sec:III-1}

The perturbation of the energy-momentum tensor $\delta t_{\mu\nu}$ is described as
\begin{equation}
 \delta t_{\mu\nu} = \delta t_{\mu\nu}^{\rm (M)} + \delta t_{\mu\nu}^{\rm (EM)},
\end{equation}
where $\delta t_{\mu\nu}^{\rm (M)}$ and $\delta t_{\mu\nu}^{\rm (EM)}$ are corresponding to the energy-momentum tensor
for the dust and the electromagnetic field, respectively.
Since the polar perturbation of fluid 4-velocity, $\delta u_\mu$, and the perturbation of density, $\delta \rho$,
are defined as
\begin{gather}
 \delta u_{\mu} = \left[\left(\tilde{\gamma}n_A + \frac{1}{2}h_{AB}u^B\right)Y^{lm},
     \tilde{\alpha}Y^{lm}_{\ \ :a}\right], \\
 \delta \rho = \tilde{\omega} \rho Y^{lm},
\end{gather}
the expansion coefficients of $\delta t_{\mu\nu}^{\rm (M)}$ in Eq. (\ref{PFP}) are
\begin{gather}
 \Delta t_{AB}^{\rm (M)} = \rho \left[\tilde{\gamma}\left(u_A n_B + n_A u_B\right)
     + \frac{1}{2}\left(h_{BC}u_A + h_{AC}u_B\right)u^C\right] + \tilde{\omega}\rho u_Au_B, \\
 \Delta t_A^{\rm (p)(M)} = \tilde{\alpha}\rho u_A, \\
 \Delta t^{2 {\rm (M)}} = \Delta t^{3 {\rm (M)}} = 0.
\end{gather}
On the other hand, as mentioned before we consider the effect of the dipole magnetic fields on
the quadrupole gravitational radiation.
In this case the non-zero expansion coefficients for $\delta t^{\rm (EM)}_{\mu\nu}$
associated with $l=2$ gravitational waves are given in the Appendix A of Paper I as
\begin{align}
 \Delta t_{\eta\eta}^{\rm (EM)} &= \kappa\left[\frac{{b_2}^2}{R^2\sin^4\chi}
               - \frac{{b_1}^2 + {b_3}^2}{R^2\sin^2\chi}\right], \\
 \Delta t_{\chi\chi}^{\rm (EM)} &= -\kappa\left[\frac{{b_2}^2}{R^2\sin^4\chi}
               + \frac{{b_1}^2 + {b_3}^2}{R^2\sin^2\chi}\right], \\
 \Delta t^{\rm (p)(EM)}_{\chi} &= \frac{\kappa \, b_1 b_2}{R^2\sin^2\chi}, \\
 \Delta t^{2 {\rm (EM)}}       &= \frac{\kappa \left({b_3}^2 - {b_1}^2 \right)}{R^2}, \\
 \Delta t^{3 {\rm (EM)}}       &= \frac{\kappa \, {b_2}^2}{R^4\sin^4\chi},
\end{align}
where $\kappa = (8\pi\sqrt{5\pi})^{-1}$.
Thus we can derive the gauge-invariant quantities for the total matter perturbations such as
\begin{align}
 T_{\eta\eta} &= -\rho k_{\eta\eta} + \omega \rho R^2 + \kappa\left[\frac{{b_2}^2}{R^2\sin^4\chi}
               - \frac{{b_1}^2 + {b_3}^2}{R^2\sin^2\chi}\right], \\
 T_{\chi\chi} &= -\kappa\left[\frac{{b_2}^2}{R^2\sin^4\chi}
               + \frac{{b_1}^2 + {b_3}^2}{R^2\sin^2\chi}\right], \\
 T_{\eta\chi} &= -\gamma \rho R^2 - \frac{1}{2} \rho k_{\eta\chi}, \\
 T_{\eta}     &= -\alpha \rho R, \\
 T_{\chi}     &= \frac{\kappa \, b_1 b_2}{R^2\sin^2\chi}, \\
 T^2          &= \frac{\kappa \left({b_3}^2 - {b_1}^2 \right)}{R^2}, \\
 T^3          &= \frac{\kappa \, {b_2}^2}{R^4\sin^4\chi},
\end{align}
where $\alpha$, $\gamma$, and $\omega$ are gauge-invariant set of fluid perturbation defined as
\begin{align}
 \alpha &\equiv \tilde{\alpha} - p^A u_A, \\
 \gamma &\equiv \tilde{\gamma} - n^A \left[p^B u_{A|B} + \frac{1}{2}u^B \left(p_{B|A} - p_{A|B}\right)\right], \\
 \omega &\equiv \tilde{\omega} - p^A(\ln \rho)_{|A}.
\end{align}

Then, with Eqs. (\ref{perturbation-01}) -- (\ref{perturbation-03}),
the evolutionary equations for the metric perturbations
on the interior region are described as
\begin{gather}
 -\partial_{\eta}^2 \zeta + \partial_{\chi}^2 \zeta - \frac{2\partial_{\eta}R}{R}\partial_{\eta}\zeta
     - \frac{2\cos\chi}{\sin\chi}\partial_{\chi}\zeta - \frac{(l-1)(l+2)}{\sin^2\chi}\zeta
     = \bar{S}_{\zeta}, \label{OS-01} \\
 \partial_{\eta}^2 k + \frac{3\partial_\eta R}{R}\partial_\eta k + \frac{\partial_\eta R}{R} \partial_\eta \zeta
     - \frac{\cos\chi}{\sin\chi}\partial_{\chi}\zeta - 2k - \left[\frac{(l-1)(l+2)}{2\sin^2\chi}+2\right]\zeta
     = \bar{S}_{k}, \label{OS-02} \\
 \partial_\eta \psi + \partial_\chi \zeta + \frac{2\partial_\eta R}{R}\psi = \bar{S}_{\psi}, \label{OS-03}
\end{gather}
where we set to be $c_s=0$ since the matter is considered as dust.
Notice that the source terms $\bar{S}_\zeta$, $\bar{S}_k$, and $\bar{S}_\psi$
are produced only by the perturbed magnetic fields
and the concrete forms are given in Appendix \ref{sec:appendix_1}.
In other words, as mentioned in \S \ref{sec:II-1-2} we can calculate the metric evolutions
apart from the matter perturbations.
At last, with Eqs. (\ref{perturbation-05}) -- (\ref{perturbation-07}), the evolutions of
matter perturbations are determined by using the variables for metric and magnetic perturbations,
which are described in Appendix \ref{sec:appendix_2}.
From the above system of equations for metric perturbations,
we can see that the variable $\zeta$ is independent of the others $k$ and $\psi$,
while the equations for variables $k$ and $\psi$ do not have such terms as
$\partial_\chi k$ and $\partial_\chi \psi$.
Thus for the interior region it is enough to calculate the evolution for only $\zeta$.
After that, in order to adopt the junction conditions (see \S\ref{sec:III-3}),
we have to calculate $k$ and $\psi$ in the vicinity of stellar surface
with the determined $\zeta$.

Considering the behavior of the metric perturbations near the stellar center,
we introduce a new variable, $\bar{\zeta}$, 
which is regular at the stellar center and defined as
$\zeta = \left(R\sin\chi\right)^{l+2}\bar{\zeta}$ \cite{Gundlach2000}.
With new variable the above perturbation equation for $\zeta$, Eq. (\ref{OS-01}),
can be rewritten as
\begin{gather}
  -\partial_\eta^2 \bar{\zeta} + \partial_\chi^2 \bar{\zeta}
          - 2(l+3)\frac{\partial_\eta R}{R}\partial_\eta\bar{\zeta}
          + 2(l+1)\frac{\cos\chi}{\sin\chi}\partial_\chi\bar{\zeta}
          - (l+2)\left[\frac{\partial_\eta^2 R}{R} + (l+3)\left(\frac{\partial_\eta R}{R}\right)^2 + l\right]
            \bar{\zeta} = \frac{\bar{S}_\zeta}{\left(R\sin\chi\right)^{l+2}}.
\end{gather}
Furthermore, in the actual numerical calculations we adopt the double null coordinates, $(u,v)$,
defined as $u=\eta-\chi$ and $v=\eta+\chi$. In these coordinates, the perturbation
equation is rewritten as
\begin{align}
 \frac{\partial^2\bar{\zeta}}{\partial u\partial v}
        &+ \frac{1}{2}\left[(l+3)\frac{\partial_\eta R}{R} + (l+1)\frac{\cos\chi}{\sin\chi}\right]
           \frac{\partial \bar{\zeta}}{\partial u} \nonumber \\
        &+ \frac{1}{2}\left[(l+3)\frac{\partial_\eta R}{R} - (l+1)\frac{\cos\chi}{\sin\chi}\right]
           \frac{\partial \bar{\zeta}}{\partial v}
         + \frac{(l+2)}{4}\left[\frac{\partial_\eta^2 R}{R} 
         + (l+3)\left(\frac{\partial_\eta R}{R}\right)^2 + l\right]\bar{\zeta}
         = -\frac{\bar{S}_\zeta}{4\left(R\sin\chi\right)^{l+2}}.
\end{align}

\subsection{Exterior region of the star}
\label{sec:III-2}

In the exterior region the master equation for perturbations can be reduced to
the well known Zerilli equation for the Zerilli function, $Z(t,r)$, which has the form
\begin{gather}
 -\partial_t^2 Z + \partial_{r_*}^2 Z - V_Z(r) Z = 0, \label{WE-out} \\
 V_Z(r) = f\left[\frac{l(l+1)}{r^2} - \frac{6M}{r^3}\frac{r^2\lambda (\lambda + 2) + 3M(r-M)}{(r\lambda + 3M)^2}\right],
\end{gather}
where $\lambda = (l+2)(l-1)/2$ and the tortoise coordinate, $r_*$, is defined as
$r_*\equiv r+2M\ln (r/2M -1)$, which leads to $\partial_{r_*}=f\partial_r$.
This variable $Z(t,r)$ is constructed with the variables $\tilde{\zeta}$ and $\tilde{k}$ as
\begin{equation}
 Z(t,r) = {\cal A}(r)\tilde{\zeta}(t,r) + {\cal B}(r)\tilde{k}(t,r) + {\cal C}(r)\partial_{r_*}\tilde{k}(t,r),
\end{equation}
where
\begin{align}
 {\cal A}(r) = \frac{2r}{\Lambda_3},\ \ 
 {\cal B}(r) = \frac{r\Lambda_1}{\Lambda_3},\ \ 
 {\cal C}(r) = \frac{-2r^2}{f\Lambda_3},
\end{align}
where $\Lambda_1 = -1+(l^2+l+1)/f$ and $\Lambda_3 = -3+(l^2+l+1)/f$ \cite{Gundlach2000}.
On the other hand, the variables of metric perturbations, $\tilde{k}$, $\tilde{\zeta}$ and $\tilde{\psi}$,
are produced with $Z$ as
\begin{align}
 \tilde{k} &= \frac{Z}{r} + \frac{2}{l(l+1)}\left[-\frac{6MZ}{r^2\Lambda_3} + \partial_{r_*}Z\right], \\
 \tilde{\zeta} &= -\frac{4r+3M}{r(2r+3M)}Z + \frac{6M}{l(l+1)}\frac{4r^2 - 7rM - 9M^2}{r^2(2r+3M)^2}Z
     - \frac{2M}{l(l+1)}\frac{5r-3M}{(r-2M)(2r+3M)}\partial_{r_*}Z + \frac{2r}{l(l+1)f}\partial_{r_*}^2 Z, \\
 \tilde{\psi}  &= -\frac{2}{l(l+1)f}\left[\left\{\frac{(l-1)(l+2)}{\Lambda_3}
     - \frac{M}{r}\right\} \partial_t Z + r \partial_t \partial_{r_*}Z\right].
\end{align}
Finally, the Zerilli equation (\ref{WE-out}) can be
rewritten in terms of the double null coordinates, $\tilde{u}=t-r_*$ 
and $\tilde{v}=t+r_*$, as
\begin{equation}
 \frac{\partial^2 Z}{\partial \tilde{u} \partial \tilde{v}} + 
      \frac{1}{4} V_Z(r) Z = 0\,.
    \label{exterior_GW}
\end{equation}
%

\subsection{Junction Conditions at the stellar surface}
\label{sec:III-3}

In order to connect the metric perturbations of interior region
with those of exterior regions, we have to impose the junction conditions
at the stellar surface, which are derived from the continuity of the induced metric and the extrinsic
curvature \cite{{Gundlach2000}}. As mentioned the above, in this paper we focus on the case that
the magnetic field would be confined inside of star, i.e., $q=0$ at stellar surface (see Eq.(\ref{q})).
In this case the junction conditions to impose are the continuity of $N$, $k$, $\zeta$, $\psi$,
$k'+8\pi\rho N$, and $\zeta'+2\mu\psi$, where the variable $N$ is determined by using $\gamma$ and $\psi$
such as $\dot{N}-\mu N = -(\gamma+\psi/2)$. These conditions are equivalent to
\begin{gather}
 \zeta = \tilde{\zeta},\ \ \ \ 
 k     = \tilde{k},\ \ \ \ 
 \psi  = \tilde{\psi}, \\
 \partial_\chi k + 8\pi \rho RN
       = \frac{\partial_\eta R}{f}\sin\chi_0 \left(\partial_t\tilde{k}\right)
       + \frac{R}{f}\cos\chi_0 \left(\partial_{r_*}\tilde{k}\right), \\
 \partial_\chi \zeta
       = \frac{\partial_\eta R}{f}\sin\chi_0 \left(\partial_t\tilde{\zeta}\right)
       + \frac{R}{f}\cos\chi_0 \left(\partial_{r_*}\tilde{\zeta}\right).
\end{gather}
With these junction conditions, the exterior variable $Z$ can be described by using the
interior variables $\zeta$, $k$, and $\psi$ \cite{{Gundlach2000}}, such as
\begin{equation}
  Z = rk + \frac{2r^4}{(l+2)(l-1)r+6M}\frac{1}{R^2}\left[\left(\frac{\cos^2\chi_0}{\sin^2\chi_0}
    + \left(\frac{\partial_\eta R}{R}\right)^2\right)\left(\zeta + k\right) + \frac{2\cos\chi_0}{\sin\chi_0}
      \frac{\partial_\eta R}{R}\psi - \frac{\cos\chi_0}{\sin\chi_0}\left(\partial_\chi k
    + 8\pi \rho RN\right) + \frac{\partial_\eta R}{R}\partial_\eta k\right].
\end{equation}

\section{Numerical procedure}
\label{sec:IV}

\begin{center}
\begin{figure}[htbp]
\includegraphics[height=7cm]{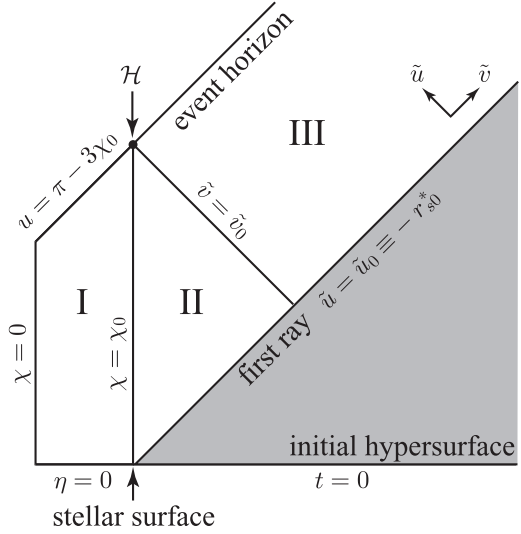}
\caption{
A schematic description of the Oppenheimer-Snyder spacetime for the collapsing 
model in characteristic coordinates.
Region I denotes the stellar interior while regions II and III correspond to
the exterior. The stellar surface, where $r=r_s$ or $\chi=\chi_0$,
is the boundary between regions I and II,
and the shaded region corresponds to the stationary region outside star.
}
\label{fig:calculation-region}
\end{figure}
\end{center}

The numerical procedure adopted in this paper is basically similar to that in Paper I.
In order to calculate the nonspherical perturbations, we divide the background spacetime
into three regions named I, II, and III (see Fig. \ref{fig:calculation-region}).
Region I denotes the stellar interior while regions II and III are corresponding to the
exterior. Region II is corresponding to the intermediate exterior region, which is
introduced to help the matching procedure at the stellar surface in numerical computation.
Region III is separated from region II via the null hypersurface defined by $\tilde{v}=\tilde{v}_0$,
which is the ingoing null ray emitted from the point where the stellar surface reaches the event
horizon, i.e., the point ${\cal H}$ in Fig. \ref{fig:calculation-region}.

In order to solve the wave equations numerically, we adopt the finite difference scheme
proposed by Hamad\'{e} and Stewart \cite{Hamade1996}, in which we use the double null coordinates
$(u,v)$ in region I and $(\tilde{u},\tilde{v})$ in regions II and III. In region I, to avoid numerical
instabilities we integrate the wave equations by using a first order finite difference scheme,
while in regions II and III the numerical integration is a second order finite difference scheme.
In region I we adopt the equally spaced grids for $(u,v)$, i.e., $\Delta u$ and $\Delta v$ are constant,
and we set to be $\Delta u=\Delta v$. With this assumption, the interval for $\eta$ and $\chi$,
$\Delta \eta$ and $\Delta \chi$, are also constant and $\Delta \eta =\Delta \chi$.
The grid points in region II are determined so that at the stellar surface they agree with the grid points produced
with the coordinates in region I. That is, $\Delta \tilde{u}$ and $\Delta \tilde{v}$ are not constant
and $\Delta \tilde{u}\ne\Delta \tilde{v}$ in region II. It is noted that with initial data sets on $\eta=0$ in region I
and on $\tilde{u}=\tilde{u}_0$ in region II, the evolutions in regions I and II can be calculated, independently of
the information in region III. After the calculation of the evolutions in regions I and II, with the data set
on the null hypersurface, $\tilde{v}=\tilde{v}_0$, and initial data set on $\tilde{u}=\tilde{u}_0$ the evolution
in region III can be calculated, where we adopt the equally spaced grids for $(\tilde{u},\tilde{v})$.
As mentioned before, hereafter we focus only on the quadrupole gravitational waves ($l=2$),
which are coupled with the dipole magnetic fields ($l=1$).

Finally it should note that in region III we calculate the time evolution for only variable $Z$,
which is subject to the Zerilli equation (\ref{exterior_GW}), while in region II, 
to make it easy to deal with the junction conditions on the stellar surface,
we also calculate the time evolutions for the variable $\tilde{\zeta}$ as well as
$Z$. The perturbation equation for $\tilde{\zeta}$ is derived from the equation (\ref{perturbation-01}),
which is described with null coordinates as
\begin{equation}
 \frac{\partial^2\tilde{\zeta}}{\partial\tilde{u}\partial\tilde{v}}
   = \frac{1}{2r}\left(1-\frac{4M}{r}\right)\left(\frac{\partial\tilde{\zeta}}{\partial\tilde{u}}
   - \frac{\partial\tilde{\zeta}}{\partial\tilde{v}}\right)
   - \frac{M}{r^3}\left(3-\frac{7M}{r}\right)\left(\tilde{\zeta}+\tilde{k}\right)
   - \frac{f}{r^2}\tilde{\zeta}.
\end{equation}
In the rest of this section, we describe the initial data, the boundary conditions at the stellar
center and at the spatial infinity, and the special treatment of the junction condition
as the stellar surface approaches the event horizon.

%

\subsection{Initial Data}
\label{sec:IV-1}

To start numerical simulations, we need to provide a data 
set on the initial hypersurface for the quantities $\bar{\zeta}$,
$\partial_u \bar{\zeta}$, and $\partial_v \bar{\zeta}$
as well as the magnetic perturbations of $b_2$ and $b_3$
for the interior region, while
$Z$, $\partial_{\tilde{u}}Z$, and $\partial_{\tilde{v}} Z$ for the exterior region. 
Outside the star, we assume that the initial perturbations are ``momentarily static'',
which is similar initial condition in \cite{Cunningham1978,SYK2007}. 
With this assumption 
the initial distribution of $Z(r;t=0)$ is determined by using the following equation; 
\begin{equation}
 -\partial_{r_*}^2 Z + V_Z(r)Z = 0,
\end{equation}
with the boundary condition at infinity as
\begin{equation}
 Z(r;t=0) \to q_2\left(\frac{2M}{r}\right)^2,
\end{equation}
where $q_2$ is a constant denoted the quadrupole moment of the star. 
Similar to \cite{Cunningham1978,SYK2007}, we assume that $q_2=2M$.
Since this solution is a static, the initial perturbation outside 
the star, $Z(r)$, does not evolve until a light signal from 
the stellar interior arrives there, i.e.,
on the gray region in Fig. \ref{fig:calculation-region} the solution $Z(r)$ will not be changed.
Thus we can use the initial data $Z(r)$ as the data set on the null hypersurface
$\tilde{u}=\tilde{u}_0$. Furthermore, with the assumption that $\partial_t Z=0$ at $t=0$,
the data for $\partial_{\tilde{u}}Z$ and $\partial_{\tilde{v}}Z$ are given as
$\partial_{\tilde{u}}Z=-(\partial_{r_*}Z)/2$ and $\partial_{\tilde{v}}Z=(\partial_{r_*}Z)/2$,
respectively.

With respect to the initial condition inside the star,
we can choose appropriate functions of magnetic distributions, $b_2(\chi)$ and $b_3(\chi)$,
where the conditions to determine the electromagnetic perturbations are
Eqs. (\ref{Maxwell1-0in}) -- (\ref{Maxwell1-2in}).
As mentioned before, since in this paper we focus only on the case that the magnetic fields
are confined inside the star, we should put the boundary conditions at $\chi=\chi_0$,
such as $b_1=b_2=b_3=0$. Then, similar to the exterior region,
if the momentarily static condition for $\bar{\zeta}$ would be assumed,
with the given initial distributions for magnetic fields,
the initial data for $\bar{\zeta}$ can be determined by
integrating the equation of
\begin{equation}
  \partial_{\chi}^2 \bar{\zeta}
         + \frac{6\cos\chi}{\sin\chi}\partial_\chi\bar{\zeta}
         - 6\bar{\zeta}
         = \frac{\bar{S}_\zeta}{\left(R\sin\chi\right)^{4}}.
  \label{initial-zeta}
\end{equation}
It should notice that for the case of the non-magnetized sphere the value of
$\partial_\chi \bar{\zeta}/\bar{\zeta}$ at the stellar surface
is independent from the central value of $\bar{\zeta}$, because the equation (\ref{initial-zeta})
does not have the source term. Thus in this case we produce the initial data of $\bar{\zeta}$
so that at the stellar surface the metric perturbation is not smooth but just continuous.
However, the effect of this non-smoothness on the emitted gravitational waves
looks like very small (see Figs. \ref{fig:waveform} and \ref{fig:test}).
Actually Cunningham, Price \& Moncrief also adopted the non-smoothness initial condition
in their calculations \cite{Cunningham1978}.
Additionally the initial data for $k$, $\partial_\chi k$, and $\psi$
at $\chi=\chi_0$ are derived from the junction conditions.
Then, as mentioned in the previous section, the variables for matter perturbations,
$\gamma(\chi)$, $\alpha(\chi)$, and $\omega(\chi)$,
can be determined by using the initial distributions for metric perturbations.
Finally we also add an assumption that $N=0$ at $\eta=0$.

\subsection{Boundary Conditions}
\label{sec:IV-2}

For the numerical integration we have to impose the boundary conditions. One is the regularity 
condition at the stellar center ($\chi=0$) and the other is the no incoming-waves condition
at the infinity. 
The regularity condition at the stellar center demands that
$\partial_\chi \bar{\zeta}=0$, which is reduced to
$\partial_u \bar{\zeta} = \partial_v \bar{\zeta}$.
With respect to the no incoming radiation condition at the infinity, we adopt the 
condition as $\partial Z/\partial\tilde{u} = 0$ (see, e.g., \cite{Hamade1996}).

\subsection{Special Treatment of the Junction Conditions near the Event Horizon}
\label{sec:IV-3}

When the stellar surface reaches the event horizon, the junction conditions
discussed earlier in \S \ref{sec:III-3} can not be used any more because the terms 
related to $f^{-1}$ diverge. 
Instead of these junction conditions, following \cite{Harada2003}, we adopt
an extrapolation for the value of $Z$ on
the junction null surface, $\tilde{v}=\tilde{v}_0$, in the vicinity of the 
point ${\cal H}$ in Fig. \ref{fig:calculation-region} as
\begin{align}
 Z &= Z^{N_{\rm max}}
     + \frac{Z^{\rm EH} - Z^{N_{\rm max}}}{r^{\rm EH} - r^{N_{\rm max}}}\left(r - r^{N_{\rm max}}\right), \\
 Z^{\rm EH} &\equiv Z^{N_{\rm max}}
     + \frac{Z^{N_{\rm max}} - Z^{N_{\rm max}-1}}{r^{N_{\rm max}} - r^{N_{\rm max}-1}}
       \left(r^{\rm EH} - r^{N_{\rm max}}\right),
\end{align}
where $Z^{n}$ and $r^{n}$ are the values of $Z$ and $r$ on 
$\tilde{v}=\tilde{v}_0$ at $n$-th time steps, while $N_{\rm max}$ denotes the 
total number of time steps in region II, and $r^{\rm EH} = 2M$.

\section{Code Tests}
\label{sec:V}

\begin{figure}[htbp]
\begin{center}
\begin{tabular}{cc}
\includegraphics[scale=0.45]{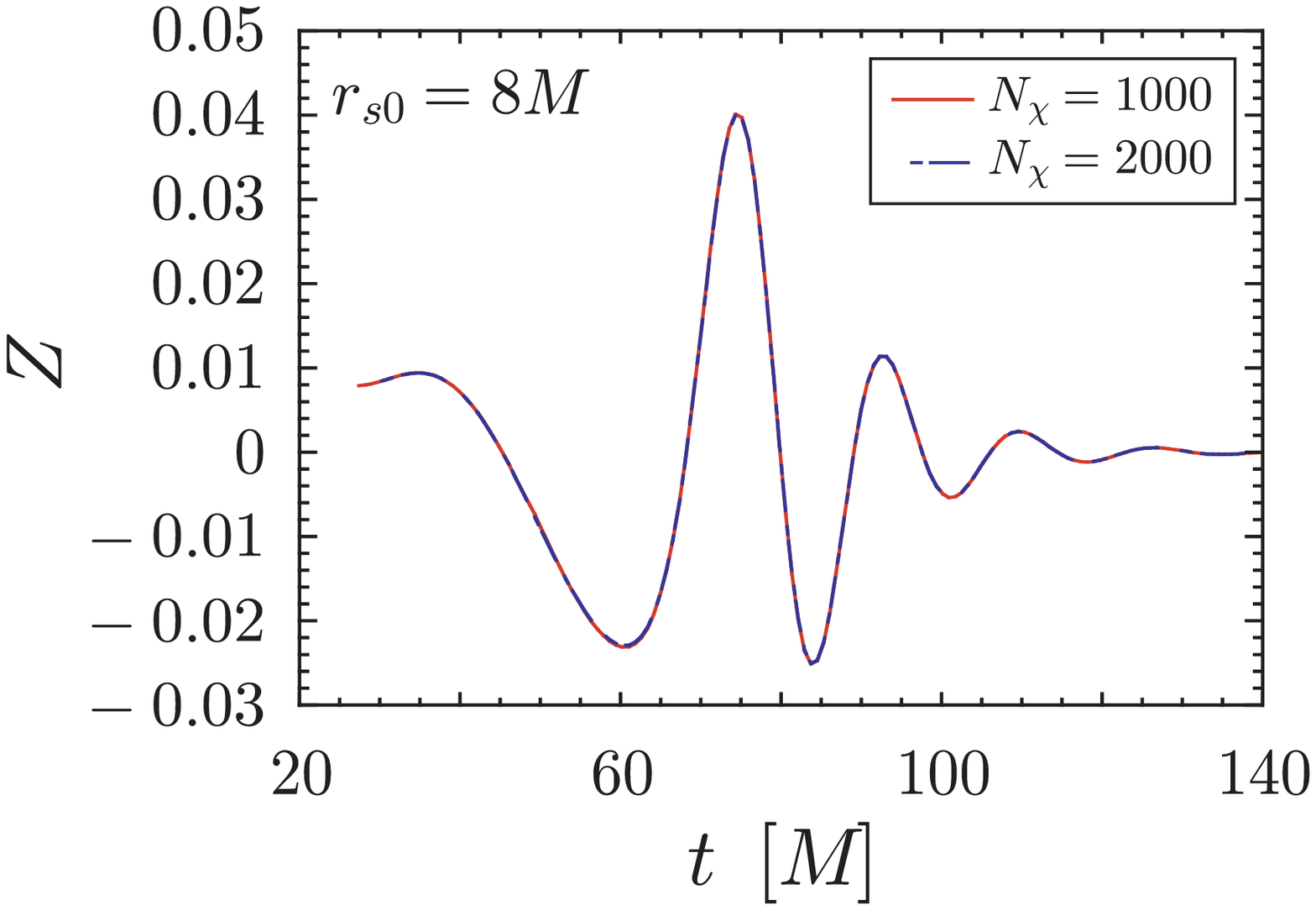} &
\includegraphics[scale=0.45]{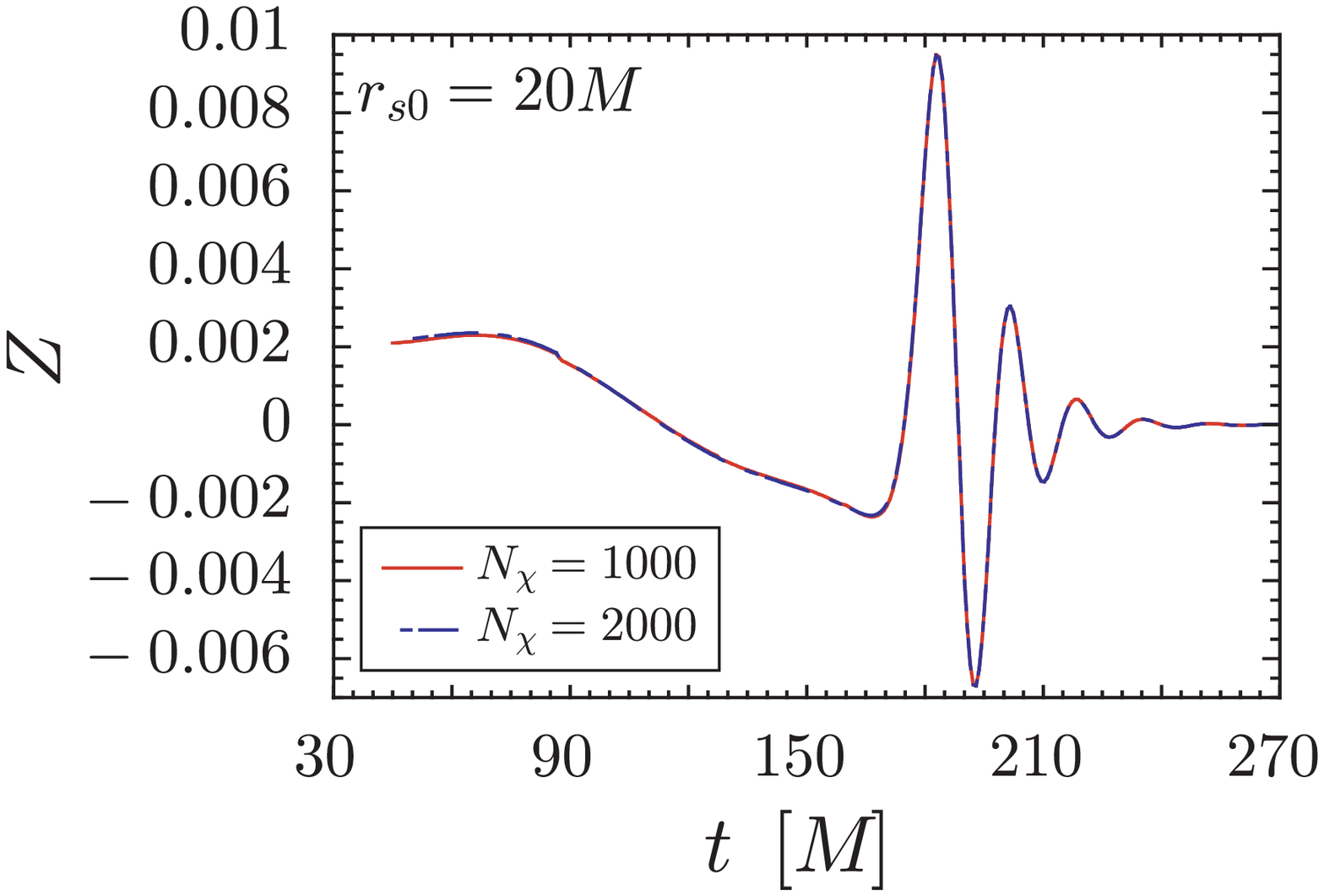} \\
\end{tabular}
\end{center}
\caption{
Waveforms of gravitational waves for $l=2$ emitted from the homogeneous
dust collapse with the initial radius $r_{s0}=8M$ (left panel) and
$r_{s0}=20M$ (right panel). The solid and broken
lines are corresponding to the results for $N_\chi = 1000$ and 2000,
respectively, where $N_\chi$ is the number of spatial grid points
inside the dust ball.
}
\label{fig:waveform}
\end{figure}

In order to verify our numerical code, we have calculated quadrupole gravitational
radiations emitted during the collapse of a nonmagnetized homogeneous dust sphere,
i.e., the gravitational waves emitted from the Oppenheimer-Snyder solution.
The number of spatial grid points inside the star, $N_\chi$, which corresponds to region I,
is chosen to be $N_\chi=1000$, because we can not see a dramatic improvement with lager
number of grid points. Actually, as shown in Fig. \ref{fig:waveform}, the waveforms
of gravitational waves emitted from the collapsing dust ball with $N_\chi=1000$
are very similar to those with $N_\chi=2000$, and the total energies of emitted gravitational waves
defined later also agree with each other within $0.626$ \% for $r_{s0}=8M$ and $1.37$ \% for $r_{s0}=20M$.
In region III, the step size for integration is given as
$\Delta \tilde{u}= (u_{\rm max}-u_0)/N_{\tilde{u}}$, where $u_{\rm max}$ is determined
with the expected maximum time for observer, $t_{\rm max}$, and the position of observer
described in tortoise coordinate, $r_{*{\rm ob}}$, as $u_{\rm max}\equiv t_{\max}-r_{*{\rm ob}}$.
In this paper we adopt that $t_{\rm max}=2000M$ and $r_{\rm ob}=r_{s0}+40M$, respectively,
where the position of observer is the same choice as the previous study by Cunningham, Price
\& Moncrief \cite{Cunningham1978}. Since the numerical code in this region is essentially the same
as that in Paper I, the number of grid points for outgoing null coordinate, $\tilde{u}$,
in region III is assumed to be $N_{\tilde{u}}=10000$ in this paper
(see  Table I in Paper I for the convergence test). Then we have only one parameter to determine
the emitted gravitational waves, i.e., the initial radius $r_{s0}$.
%
%
%

\begin{figure}[htbp]
\begin{center}
\begin{tabular}{cc}
\includegraphics[scale=0.45]{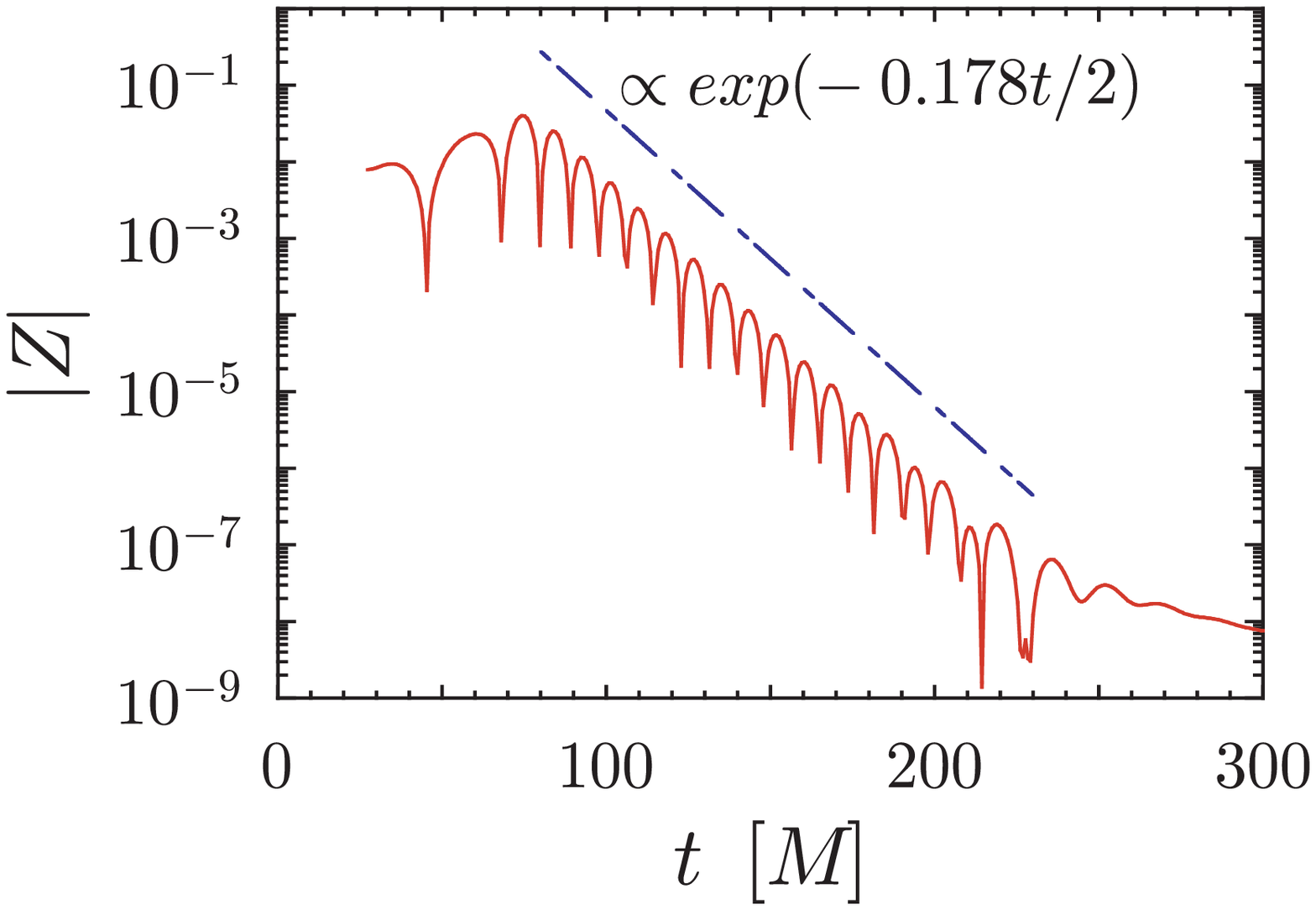} &
\includegraphics[scale=0.45]{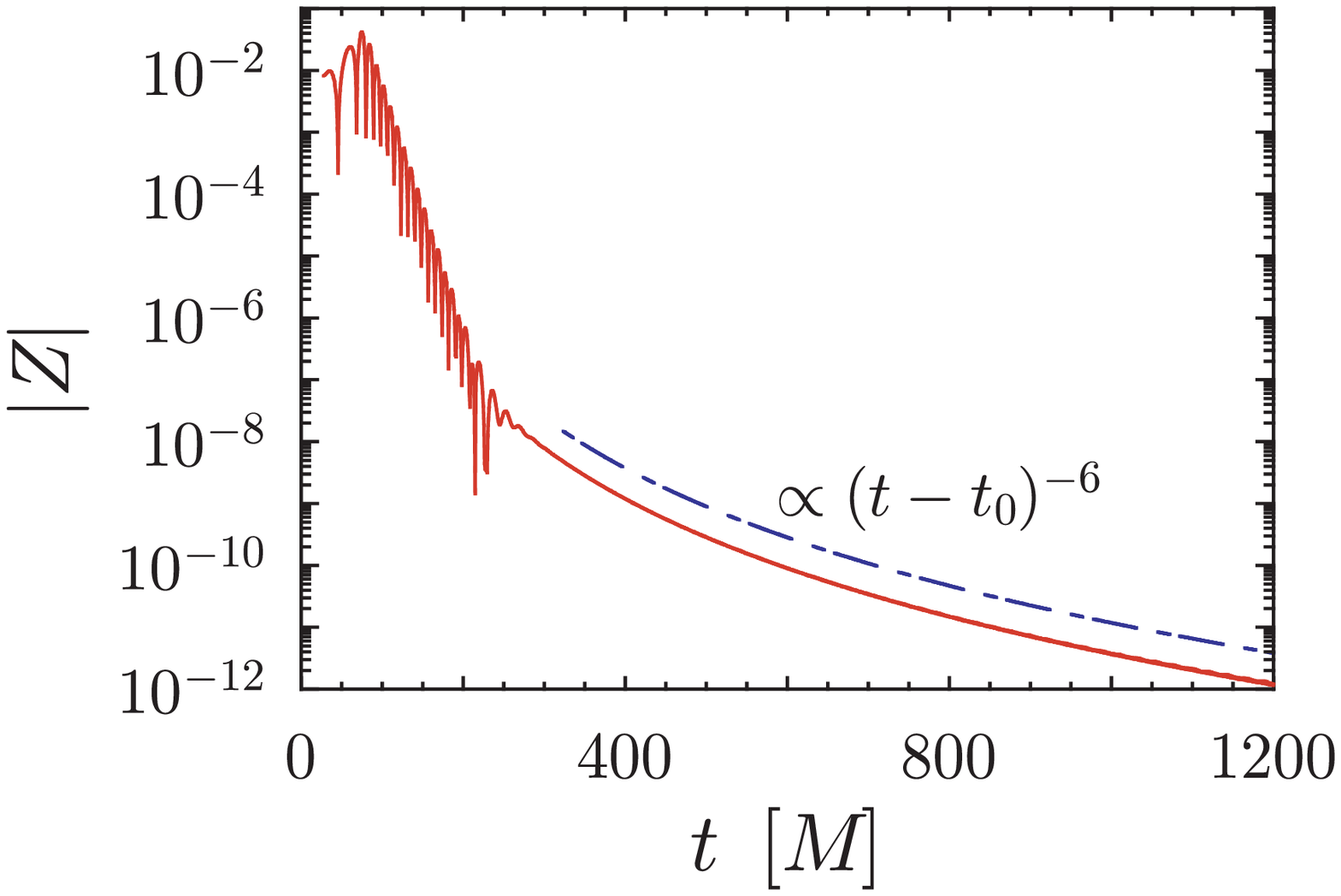} \\
\end{tabular}
\end{center}
\caption{Waveforms of the quadrupole gravitational radiation emitted during 
the collapse of the non-magnetized homogeneous dust,  as function of  time.
The initial radius of the dust sphere is set to $r_{s0}=8M$
while the fiducial observer is set at $r_{\rm ob}=r_{s0}+40M$. 
In the right panel the late-time is compared with its theoretical value $(t-t_0)^{-(2l+2)}$,
where $t_0$ is the time when the observer receives the first signal emitted from the stellar surface.
}
\label{fig:test}
\end{figure}

As noticed in \cite{Cunningham1978}, the emitted gravitational waves are characterized
by the quasi-normal ringing oscillation and subsequent power-law tail.
In Fig. \ref{fig:test}, we show the waveform of the gravitational waves for $l=2$
emitted during the collapse of the homogeneous dust ball, where the left and right
panels are focused on the quasi-normal ringing and on the power-law tail, respectively.
The fundamental frequency of the quasi-normal ringing have been calculated by Chandrasekhar
\& Detweiler \cite{Chandrasekhar1975}, such as $2M\omega=0.74734+0.17792i$. On the other hand,
our numerical results show that the oscillation frequency is $2M\omega =0.737$,
which agrees well with the previous value with only $1.3\%$ error, while the damping rate
also consorts with the theoretical value (see the left panel of Fig. \ref{fig:test}).
As for the late-time tail, in the right panel of Fig. \ref{fig:test}, we find that the
amplitude of gravitational wave decays as $(t-t_0)^{-6}$, where $t_0$ is the time when
the observer receives the first signal emitted from the stellar surface, i.e.,
$t_0\equiv r_{*\rm ob} - r_{*s0}$. This result is in good agreement with the analytical
estimate by Price \cite{Price1972}, that is $(t-t_0)^{-(2l+2)}$. Through these estimations for
the frequency of quasi-ringing and the late-time tail, we believe that our numerical code
is possible to derive the gravitational waves with high accuracy.
%
%
%
%
%

\begin{center}
\begin{figure}[htbp]
\includegraphics[height=6cm]{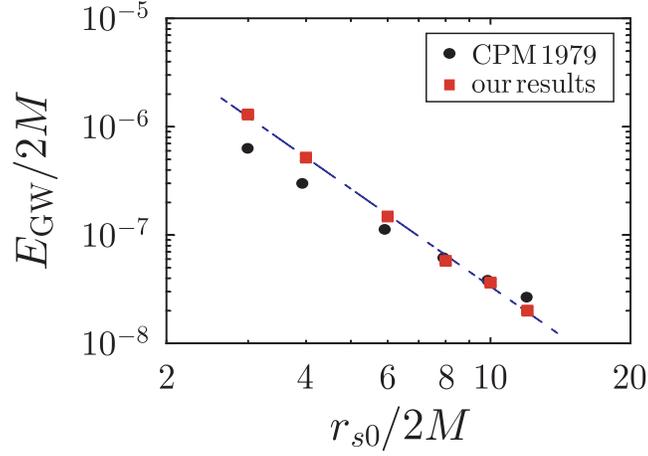}
\caption{
Total energies emitted in gravitational waves from the homogeneous
dust collapse without magnetic field as a function of the initial
stellar radius, where $r_{s0}=6M$, $8M$, $12M$, $16M$, $20M$, and
$24M$. The filled circles correspond to the results by Cunningham,
Price, and Moncrief \cite{Cunningham1978}, while the squares are
corresponding to our numerical results. The broken line denotes
the empirical formula derived from our results.
}
\label{fig:energy}
\end{figure}
\end{center}

At the end in this section, we compare the total energy emitted during the collapse with the
previous results by Cunningham, Price, \& Moncrief \cite{Cunningham1978} (CPM1979).
It is worth to notice that the variables inside the star adopted in CPM1979 are different
from those in the equation system for the gauge-invariant formalism proposed by Gerlach \& Sengupta
\cite{Gerlach1979} and Gundlach \& Mart\'{\i}n-Garc\'{\i}a \cite{Gundlach2000}.
The total emitted energy, $E_{\rm GW}$, is estimated by integrating the luminosity of
gravitational waves, $L_{\rm GW}$, with respect to time, where the luminosity is defined as
\begin{equation}
 L_{\rm GW} = \frac{1}{384\pi}\left({Z}_{,t}\right)^2, \label{Lgw}
\end{equation}
for the $l=2$ gravitational waves (e.g., \cite{Cunningham1978}).
In Fig. \ref{fig:energy} we show the total emitted energy of gravitational waves
as a function of the initial stellar radius $r_{s0}$, where for comparison
we also plot the result of CPM1979. It notes that in this figure
we adopt the normalization for the quadrupole moment, $q_2$, so that $q_2=2M$ as mentioned before.
This figure shows that there are small difference between our results and those obtained in CPM1979.
The main reason for this difference could be the difference how to choose the variables inside
the star. Additionally as we noticed in Paper I, the difference of the accuracy in numerical code
might be also added in the reason. Anyway we can observe that the total emitted energy systematically
decreases as the initial stellar radius increases and that the emitted energy is very similar to
that of CPM1979. Furthermore, with our variables inside the star and with our initial data,
we derive the empirical formula for the total emitted energy as
\begin{equation}
 \frac{E_{\rm GW}}{2M} = 3.37\times 10^{-5}\times \left(\frac{r_{s0}}{2M}\right)^{-3},
 \label{eq:empirical}
\end{equation}
which is also plotted in Fig. \ref{fig:energy}.
%
%
%
%

\section{Gravitational Radiations from the OS solution}
\label{sec:VI}

In order to calculate the gravitational waves emitted from the collapsing phase of a magnetized dust
sphere, we have to provide the initial distribution of the magnetic field. In other words, one needs
to set up the functional forms of $b_2$ and $b_3$ on the hypersurface $\eta=0$.
The initial distributions can be determined as the following two conditions are satisfied;
(a) the regularity condition at the stellar center and (b) the junction condition at the stellar surface.
Since we made assumptions in this paper that the magnetic field is confined inside the star and 
then the value of $q$ becomes zero at stellar surface, the conditions at the stellar surface can be
described as $b_1(\chi_0)=b_2(\chi_0)=b_3(\chi_0)=0$.
Now we introduce two new variables, $\bar{b}_2$ and $\bar{b}_3$, such as
\begin{align}
 b_2(\chi) &= {\cal B}_2\sin^2\chi\, \bar{b}_2(\chi), \\ 
 b_3(\chi) &= {\cal B}_3\sin^3\chi\, \bar{b}_3(\chi), 
\end{align}
where ${\cal B}_2$ ${\cal B}_3$ are arbitrary constants related to the strength of the magnetic field.
With analytic functions $\bar{b}_2$ and $\bar{b}_3$, the regularity condition at the stellar center for
the magnetic field is automatically satisfied.
Since the geometry of the magnetic field when the collapse sets in is practically unknown,
in this paper we adopt the following two types of the initial distributions for the magnetic field;
\begin{align}
 {\rm (I) : }&\ \bar{b}_2(\chi) = \bar{b}_3(\chi)
                = 1 - 2\left(\frac{\chi}{\chi_0}\right)^2 + \left(\frac{\chi}{\chi_0}\right)^4, \\
 {\rm (II) : }&\ \bar{b}_2(\chi) = \bar{b}_3(\chi)
                 = 16\left(\frac{\chi}{\chi_0}\right)^4
            \left[1 - 2\left(\frac{\chi}{\chi_0}\right)^2 + \left(\frac{\chi}{\chi_0}\right)^4\right],
\end{align}
where the maximum value of $\bar{b}_2$ and $\bar{b}_3$ are chosen to be one
in the range of $0\le \chi\le\chi_0$. For the first profile (I) the magnetic field is stronger in the
center of the sphere, while for the second profile (II) the field becomes stronger in the outer region.
Additionally it notes that for both profiles the value of $b_1(\chi)$,
defined as $b_1(\chi)=-\partial_\chi b_2(\chi)/2$, becomes zero at the stellar surface,
i.e., as mentioned before the value of $q$ is zero at $\chi=\chi_0$.
With these magnetic profiles, we found that the allowed values for ${\cal B}_2$ and ${\cal B}_3$ have to
be in a part of the range of ${\cal B}_2 < {\cal B}_3$,
in order to produce the initial data set so that
the inner metric perturbation should be smoothly connected to the stationary solution in the outer region
at the stellar surface.
So, in what follows, we consider the two cases for magnetic field;
one is that only toroidal magnetic component exists, i.e., ${\cal B}_2=0$, and
second is that the poloidal magnetic component also exists as well as toroidal
one, where those are satisfied the condition that ${\cal B}_2 < {\cal B}_3$.


\subsection{Toroidal Magnetic Field}
\label{sec:VI-1}

\begin{figure}[htbp]
\begin{center}
\begin{tabular}{cc}
\includegraphics[scale=0.45]{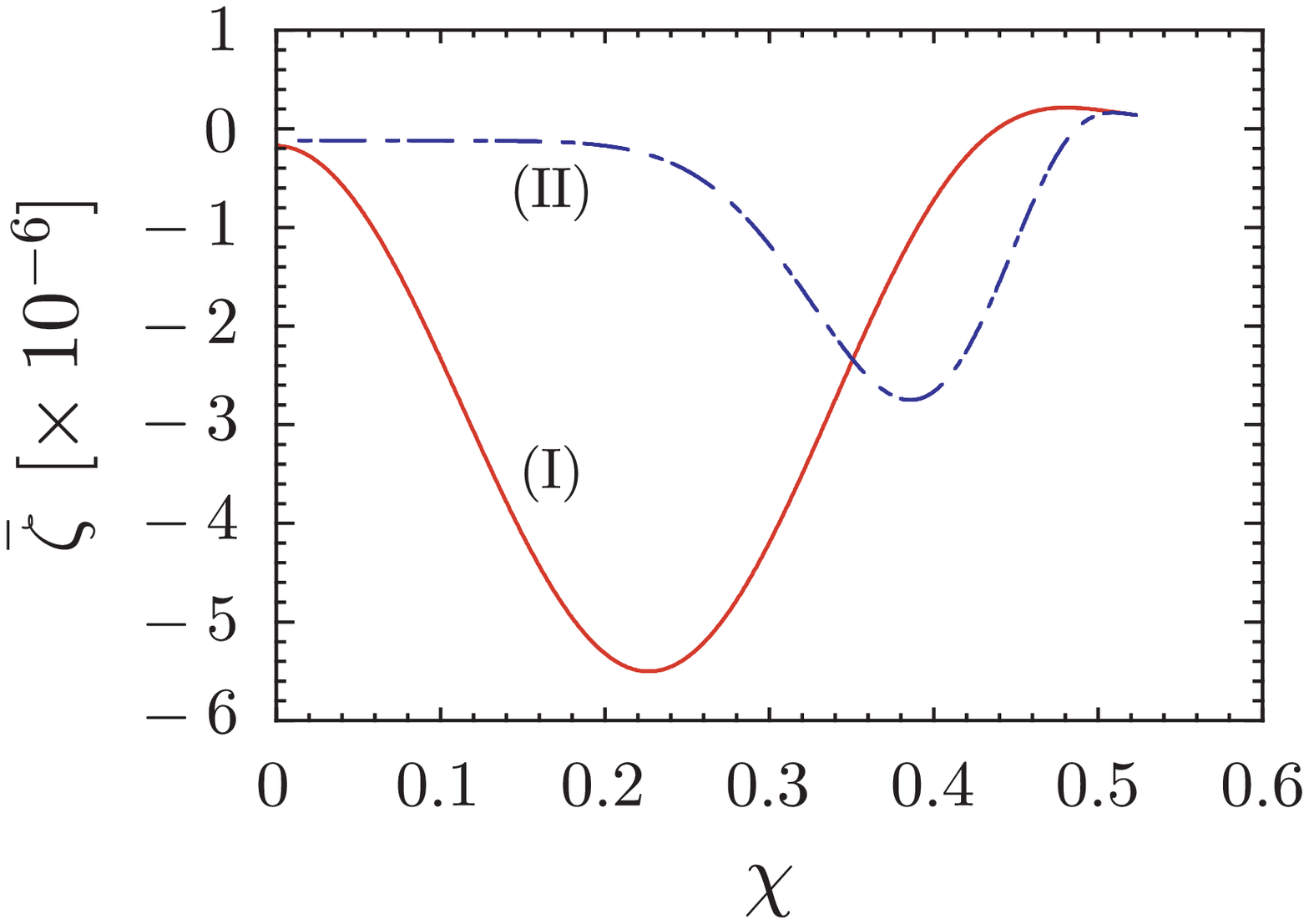} &
\includegraphics[scale=0.45]{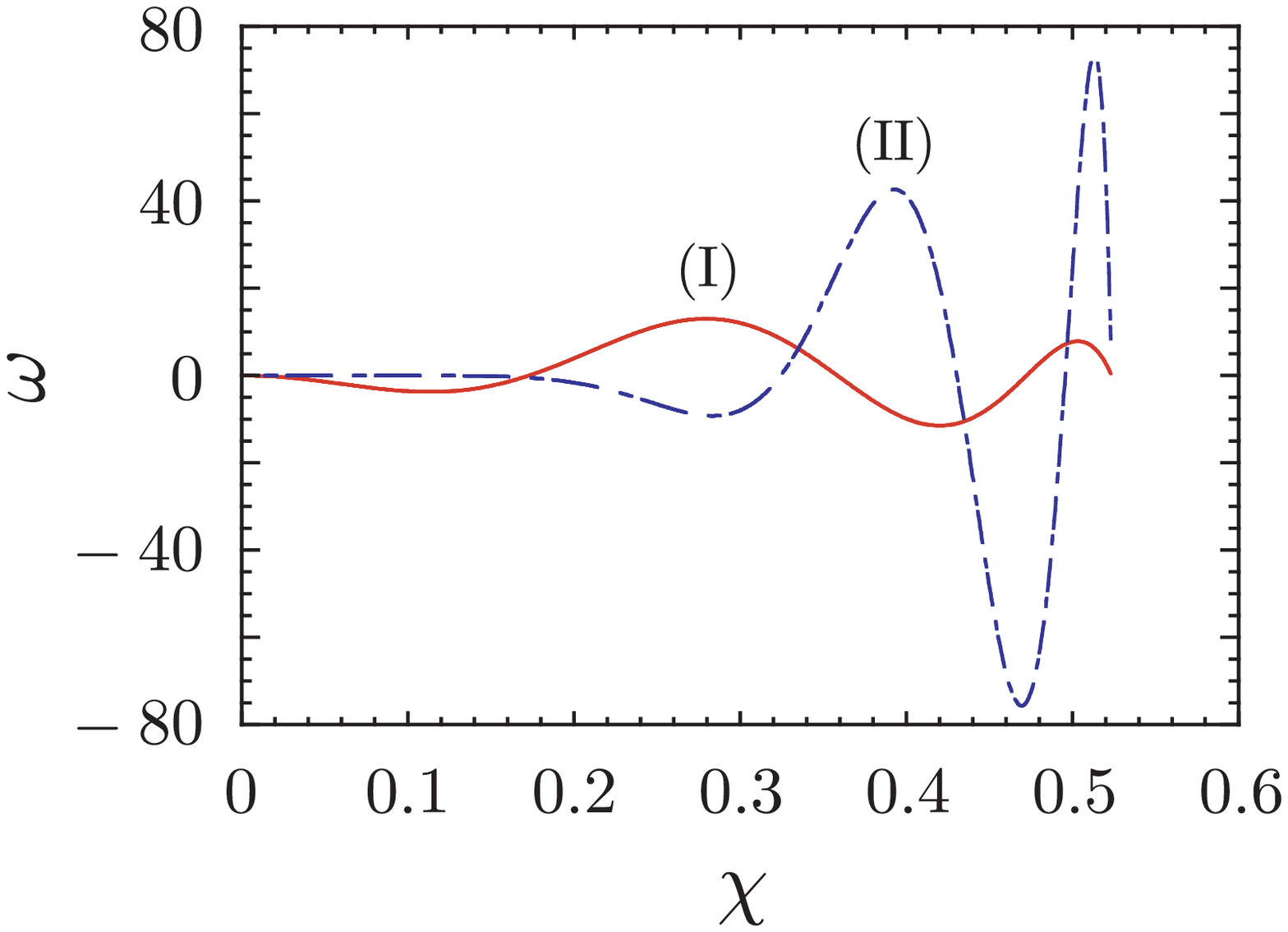} \\
\end{tabular}
\end{center}
\caption{
The initial distribution of inner metric perturbation, $\bar{\zeta}(\chi)$, on the left panel and
the initial density perturbation, $\omega(\chi)$, on the right panel,
where the initial radius is adopted that $r_{s0}=8M$.
}
\label{fig:initial-toroidal}
\end{figure}

\begin{figure}[htbp]
\begin{center}
\begin{tabular}{cc}
\includegraphics[scale=0.45]{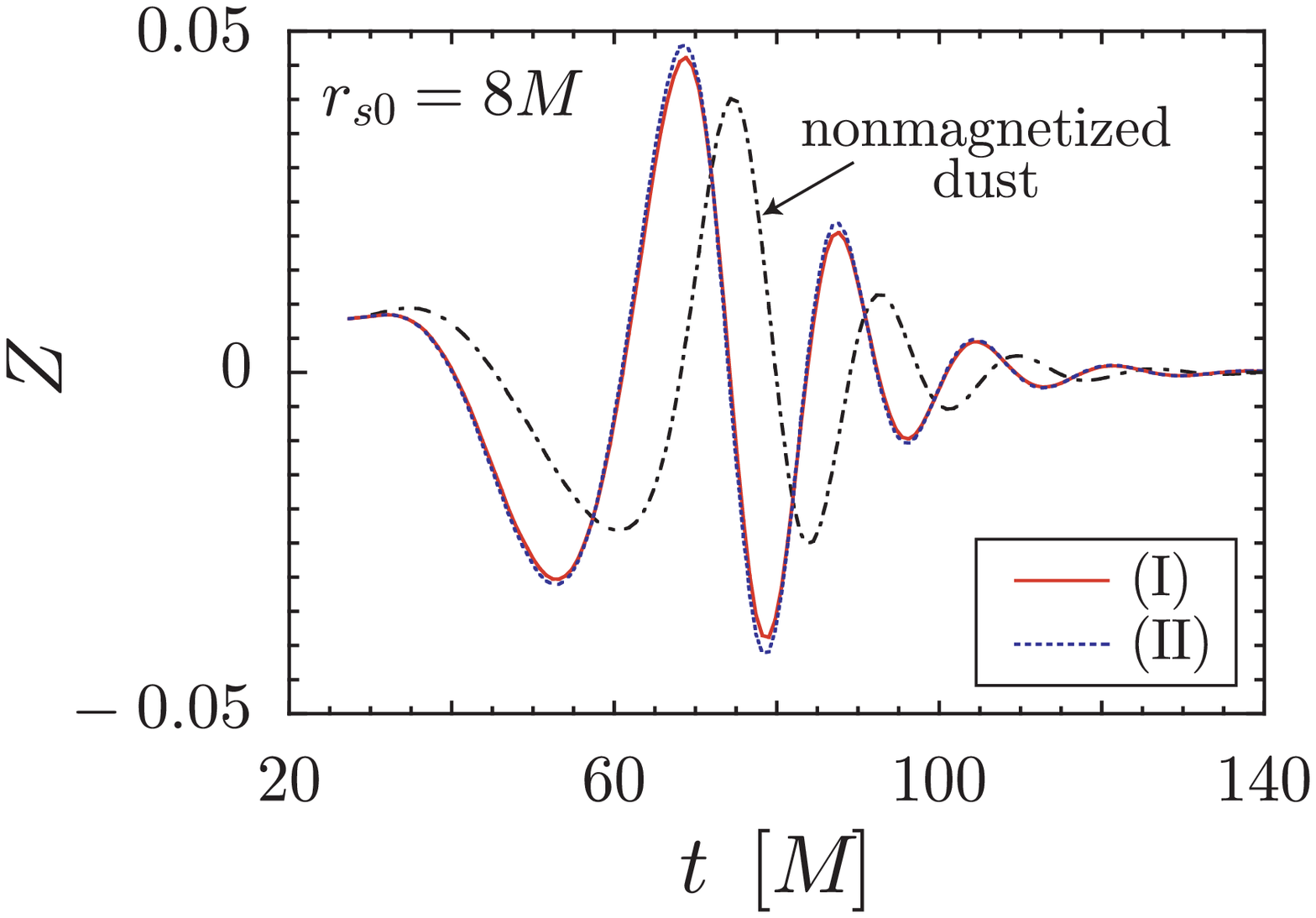} &
\includegraphics[scale=0.45]{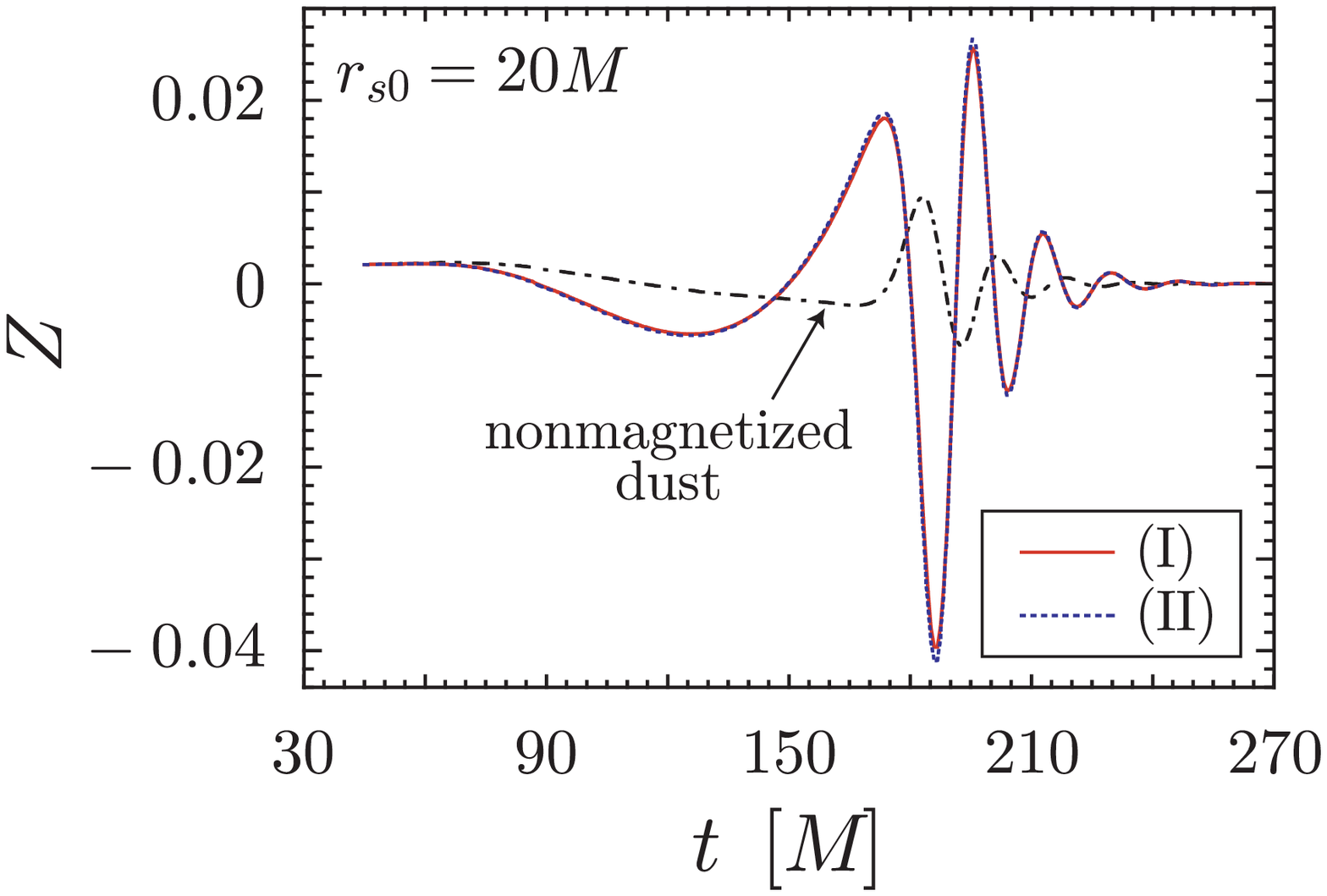} \\
\end{tabular}
\end{center}
\caption{
Waveforms of gravitational waves for $l=2$ emitted from the homogeneous
magnetized dust collapse with the initial radius $r_{s0}=8M$ on the left panel
and with $r_{s0}=20M$ on the right panel.
The solid and dotted lines are corresponding to the results with the magnetic
profile (I) and (II), respectively, while for comparison the result
for the non-magnetized case is also denoted with the dot-dash line. 
}
\label{fig:waveform-toroidal}
\end{figure}

\begin{center}
\begin{figure}[htbp]
\includegraphics[height=6cm]{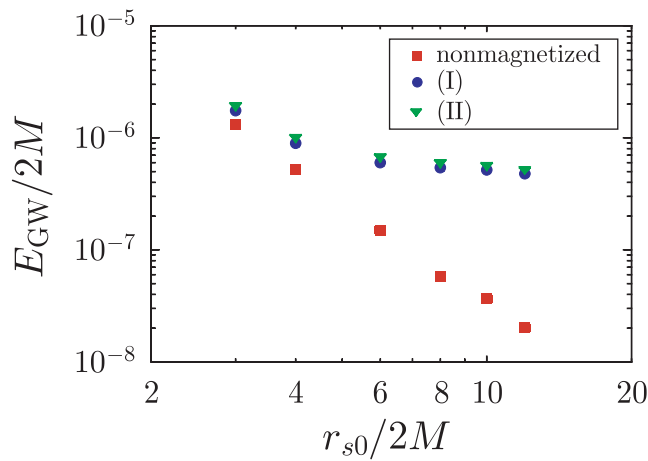}
\caption{
Total energies emitted in gravitational waves from the homogeneous
magnetized dust collapse as a function of the initial
stellar radius, where $r_{s0}=6M$, $8M$, $12M$, $16M$, $20M$, and
$24M$. The filled circles and triangles correspond to the calculated results
with magnetic profile (I) and (II), respectively, while the squares are
corresponding to the total energy emitted from the non-magnetized dust
collapse.
}
\label{fig:energy-toroidal}
\end{figure}
\end{center}

First we consider the case that only toroidal magnetic component exists, i.e., ${\cal B}_2=0$.
In this case, the source term in the equation (\ref{initial-zeta}) to determine the initial distribution,
$\bar{\zeta}$, is proportional to ${{\cal B}_3}^2$.
The value of ${\cal B}_3$ is determined so that the initial inner metric perturbation
should be smoothly connected at the stellar surface to the stationary solution
for exterior region. Then we can get the distributions for the initial inner metric perturbation,
$\bar{\zeta}(\chi)$, and for the initial density perturbation, $\omega(\chi)$, which are shown
in Fig. \ref{fig:initial-toroidal} with the two different magnetic profiles (I) and (II).
In this figure the initial stellar radius is set to be $r_{s0}=8M$,
but the functional forms of $\bar{\zeta}(\chi)$ and $\omega(\chi)$ with the different initial stellar
radii are very similar to that with $r_{s0}=8M$.
From this figure, we can see that the initial distributions of
$\bar{\zeta}$ and $\omega$ depend strongly on the magnetic profiles
even if the initial metric perturbations for exterior region are adopted the same
as the stationary solution with $q_2=2M$.
%
%
%
On the other hand, Fig. \ref{fig:waveform-toroidal} shows the waveforms of the emitted gravitational
waves with these initial perturbations, where the left and right panels are corresponding to the results
with the initial radius $r_{s0}=8M$ and $20M$, respectively. For comparison, the waveforms for
the non-magnetized dust collapse are also plotted. The first observation of Fig. \ref{fig:waveform-toroidal}
is that the waveforms of the emitted gravitational waves are almost independent from the magnetic profiles
in spite of the difference of initial perturbations inside the star.
That is, if the initial dust sphere consists only of the toroidal magnetic component,
it might be difficult to distinguish the interior magnetic profile
by using the direct detection of the waveform of the emitted gravitational waves.
Additionally we can observe the difference between the waveforms of gravitational waves
with the toroidal magnetic field and without magnetic field.
With smaller initial radius, the shape of waveform is similar to that for the nonmagnetized case,
still we can see the effect of the existence of magnetic field, i.e., the quasi normal ringing can be seen
earlier and the amplitude is also enhanced a little due to the magnetic effect.
While, with large initial radius, it is possible to watch the obvious influence of magnetic field
on the waveform of emitted gravitational waves, where the amplitude of waveform grows large and
the maximum value of gravitational wave becomes a negative. In other words, with large initial radius,
the waveform before the quasi normal ringing would be observed can be changed remarkably.
The reason for this could be thought that with large initial radius it takes longer time until the stellar surface
reaches to the event horizon and then the inner magnetic field can affect on the metric perturbations with longer time.
%
%
%
In particular, the dependence of magnetic effect on the initial radius can be seen clearly
in the total energy of emitted gravitational waves.
Fig. \ref{fig:energy-toroidal} shows the total energy as a function of the initial stellar radius
with the circles for magnetic profile (I) and with the triangles for magnetic profile (II), where
for comparison the total energies for the nonmagnetized dust case are also shown with the squares.
It is found from this figure that with large initial radius, due to the magnetic effect
the total energy of emitted gravitational wave becomes much larger than
that for the nonmagnetized dust collapse. Then the dependence of
the total energy for the dust collapse with toroidal magnetic field
on the initial stellar radius, is quite different from the empirical formula (\ref{eq:empirical})
for the nonmagnetized dust collapse.

%
%
%
%

\subsection{Poloidal and Toroidal Magnetic Fields}
\label{sec:VI-2}

Next we consider the magnetic field, which consists of the poloidal and toroidal components.
In this case we can introduce the new parameter, $\beta$, defined as $\beta={\cal B}_2/{\cal B}_3$,
and if we choose the value of $\beta$ the initial inner metric perturbations are determined as
those should be smoothly connected to the outer stationary solution.
It notes that the case for $\beta=0$ corresponds to the dust model, which only toroidal magnetic
component exists shown in the previous subsection.
Table \ref{tab:beta} shows the allowed maximum values of $\beta$ with different combinations of
magnetic profiles for the poloidal and toroidal components and with different initial stellar radii.
From this table, it can be seen that with large initial radius it becomes more difficult to
produce a magnetized dust model with large value of $\beta$, and that the maximum values of $\beta$
depends strongly on the inner magnetic profiles. 
%
%
\begin{table}[htbp]
\begin{center}
\leavevmode
\caption{ Allowed maximum values of $\beta$ with the different combinations of magnetic profiles
for the poloidal and toroidal components and with different initial stellar radii.
As mentioned before, $b_2$ and $b_3$ are associated
with the poloidal and toroidal magnetic components, respectively.
}
\begin{tabular}{cc cc |c cc cc cc cc cc c}
\hline\hline
\multicolumn{4}{c}{profile} & \multicolumn{12}{|c}{$r_{s0}$} \\
\hline
 $b_2$ & & $b_3$ & & & $6M$ & & $8M$ & & $12M$ & & $16M$ & & $20M$ & & $24M$ \\
\hline
 (I)  & & (I)  & & & $0.073$ & & $0.054$ & & $0.035$  & & $0.026$  & & $0.021$  & & $0.017$   \\
 (I)  & & (II) & & & $0.26$  & & $0.20$  & & $0.13$   & & $0.10$   & & $0.080$  & & $0.067$   \\
 (II) & & (I)  & & & $0.020$ & & $0.014$ & & $0.0094$ & & $0.0070$ & & $0.0055$ & & $0.0046$  \\
 (II) & & (II) & & & $0.072$ & & $0.053$ & & $0.035$  & & $0.026$  & & $0.021$  & & $0.017$  \\
\hline\hline
\end{tabular}
\label{tab:beta}
\end{center}
\end{table}
%
%

\begin{figure}[htbp]
\begin{center}
\begin{tabular}{cc}
\includegraphics[scale=0.45]{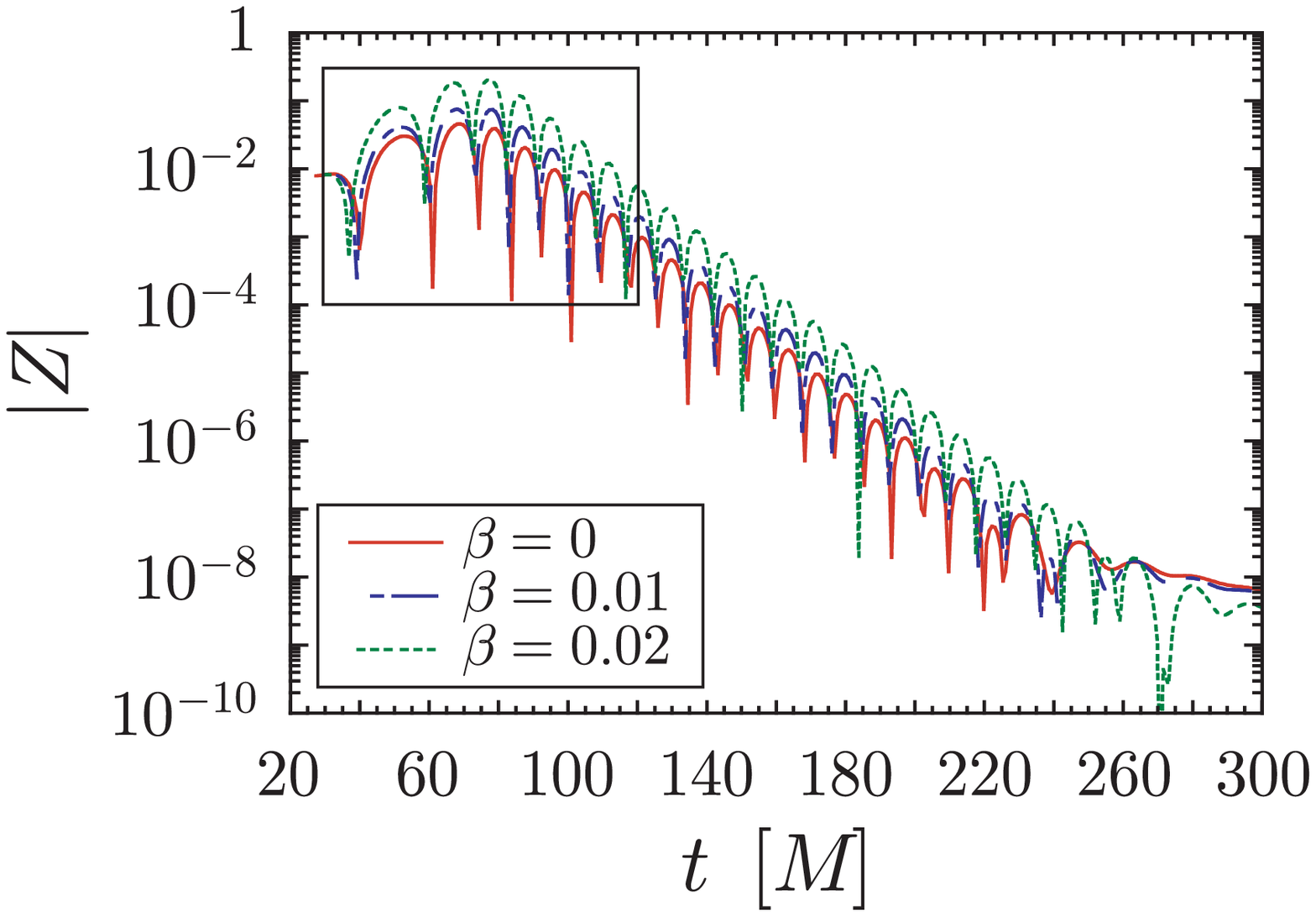} &
\includegraphics[scale=0.45]{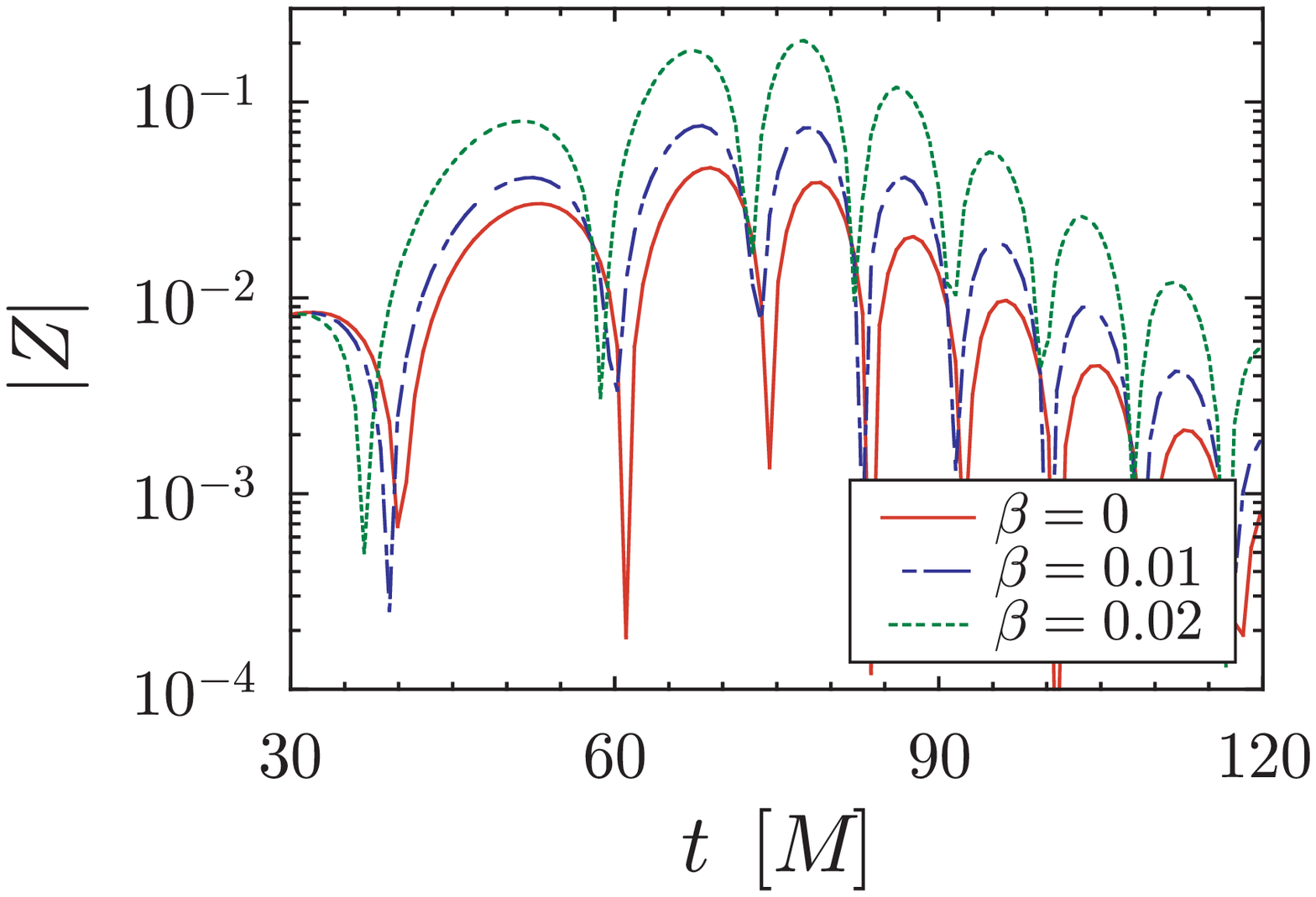} \\
\end{tabular}
\end{center}
\caption{
Waveforms of gravitational waves emitted from the homogeneous magnetized dust collapse with
the initial radius $r_{s0}=8M$ on the left panel, where three different lines are corresponding
to the dust models with different magnetic ratio, $\beta$, and the magnetic profile (I) is adopted
for both poloidal and toroidal components. It notes that the line for $\beta=0$
is the result of collapse without poloidal magnetic component.
The right panel is the magnification of the region encompassed by the square in the left panel.
}
\label{fig:waveform-T1P1-R08}
\end{figure}

\begin{figure}[htbp]
\begin{center}
\begin{tabular}{cc}
\includegraphics[scale=0.45]{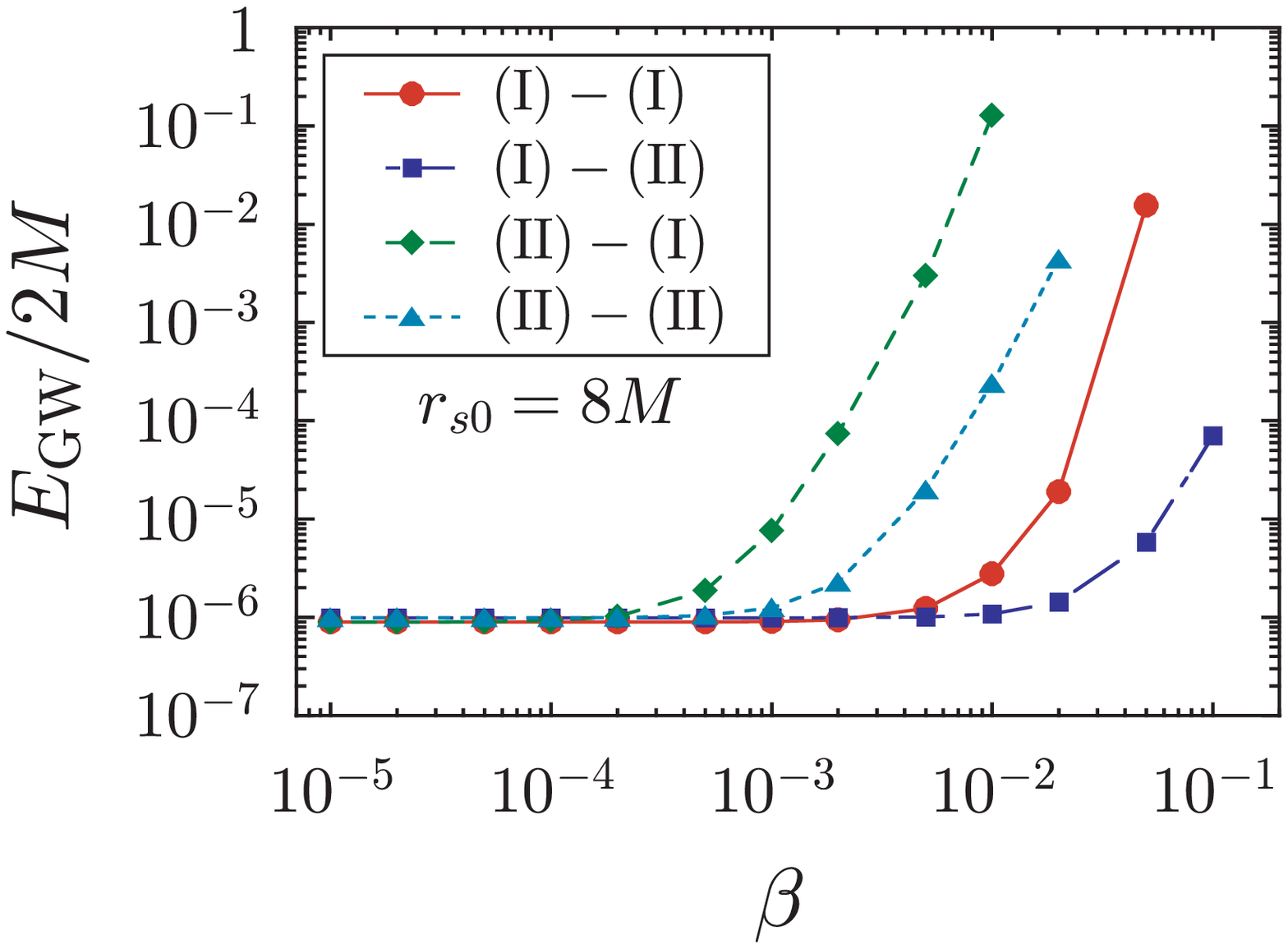} &
\includegraphics[scale=0.45]{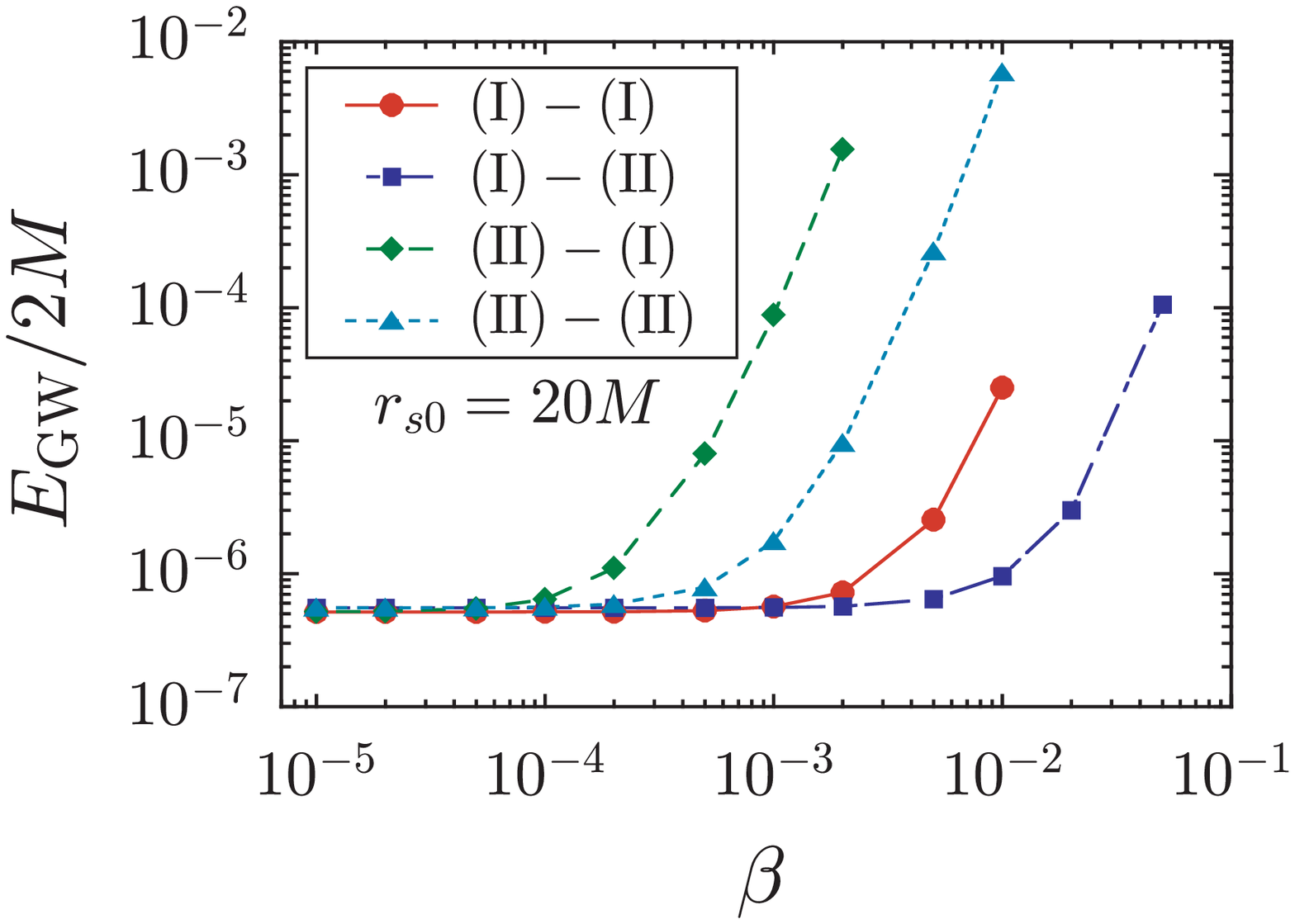} \\
\end{tabular}
\end{center}
\caption{
The total energies emitted in gravitational waves
from the homogeneous dust collapse with magnetic field as a function of the magnetic ratio, $\beta$,
where the left and right panels correspond to the results for the initial radius $r_{s0}=8M$
and for $r_{s0}=20M$. In the figure, the different lines are corresponding to
the different combinations of magnetic profiles for the poloidal and toroidal components.
In the explanatory notes, for example, ``(I)--(II)" denotes that the magnetic profiles (I)
and (II) are adopted for the magnetic variables, $b_2$ and $b_3$, respectively.
}
\label{fig:energy-PT}
\end{figure}

Similar to the collapse of magnetized dust with only toroidal component, the magnetic effects can be seen
in the waveforms of gravitational waves a little. Fig. \ref{fig:waveform-T1P1-R08} shows waveforms
for the collapse of magnetized dust with $r_{s0}=8M$ and with several values of $\beta$,
where the magnetic profile (I) is adopted for the poloidal and toroidal components.
From this figure it is found that the emitted gravitational waves are basically
characterized by the quasi-normal ringing as well as the case of nonmagnetized dust collapse.
While, we can also see the specific magnetic effects in the waveforms, where as the value of $\beta$
becomes larger, the amplitude of gravitational waves is enhanced and its maximum value changes
from a positive to a negative. These are similar features to the case of dust collapse with only toroidal
component with large initial radius. In other words, with large value of magnetic ratio,
even with small initial radius, we can see the magnetic effect in the waveform before the quasi-normal
ringing would be observed. This tendency holds for the magnetized dust collapse with different magnetic
profiles. As a result, with large value of magnetic ratio, the total energy of gravitational waves grows.
The total energies for $r_{s0}=8M$ and $20M$ are plotted in Fig. \ref{fig:energy-PT} as a function of
magnetic ratio, where the different lines are corresponding to the different combinations of
magnetic profiles for the poloidal and toroidal components. In the explanatory notes of this figure,
for example, ``(I)--(II)" denotes that the magnetic profiles (I) and (II) are adopted for the magnetic
variables, $b_2$ and $b_3$, respectively. This figure tells us that the total energies emitted in gravitational
waves from the magnetized dust collapse depend strongly on the magnetic ratio and the inner magnetic profiles.
This sensitivity, as well as the change of waveforms due to the existence of magnetic fields,
could be important to extract some information of inner magnetic profiles of progenitor
from the direct observation of gravitational waves during the black hole formation after the stellar collapse.

%
%
%
%
%
%
%
%

\section{Conclusion}
\label{sec:VII}

In this article, with the gauge-invariant perturbation theory we have studied the dependence of
the stellar magnetic fields on the polar gravitational waves during the collapse of a homogeneous
dust sphere. It should be emphasized that this is first calculation of emitted polar gravitational
waves on the dynamical background spacetime with the covariant gauge-invariant formalism
on the spherically symmetric spacetime and the coordinate-independent matching conditions at stellar surface,
which is devised by Gundlach and Mart\'{\i}n-Garc\'{\i}a \cite{Gundlach2000}. So far, such calculations
could not be done due to the difficulty to treat the boundary conditions at stellar surface.
While, in order to solve this difficulty, we evolve not only Zerille function, $Z$, but also metric perturbation,
$\tilde{\zeta}$, in the intermediate exterior region (region II) and the calculation of emitted gravitational
waves is successful.

With this nemerical code, we consider the magnetic effects on the polar gravitational waves from the
Oppenheimer-Snyder solution describing collapsing dust, where the magnetic fields are introduced as a second
order perturbation term. Even if the initial magnetic perturbations are small, as the collapse proceeds
they could get amplified and become significant because of the conservation of magnetic flux.
In particular, similar to Paper I, we have assumed that the magnetic field is axisymmetric, where the dipole
magnetic field perturbations are the ones that couple to the quadrupole polar perturbations of the gravitational
field. Additionally, we assumed momentarily static initial data and we have not taken into account
the influence of the exterior magnetic field in the propagating gravitational waves.

Through the investigation, it is found that there exists an evidence of the strong influence of the magnetic field
in the gravitational wave luminosity during the collapse.
Depending on the initial profile of the magnetic field and its ratio between the poloidal and toroidal components,
the energy outcome can be easily up to a few order higher than what we get from the nonmagnetized collapse.
In addition, it is possible to observe an important change before the quasi-normal ringing is detected, which
is induced by the presence of the magnetic field. These magnetic effects can be seen the collapsing model with
large inital radius and with large magnetic ratio between the  poloidal and toroidal components, since for a large
initial radius the time needed for the black hole formation is longer and then the magnetic field acts for longer
time on the collapsing fluid. It notes that the magnetic effects on the polar gravitational waves
are different from those on the axial ones, i.e., the axial gravitational waves are independent from the magnetic
ratio while they depend only on the magnetic strenght such as the value of ${\cal B}_2\times{\cal B}_3$.
Such magnetic effects could be helpful to extract some information of inner
magnetic profiles of progenitor form the detection of gravitational waves radiated from the black hole formation
after the stellar collapse.

At the end, we believe that although this study might be considered as a ``toy problem'' it has most of the
ingredients needed in emphasizing the importance of the magnetic fields in the study of the gravitational wave output
during the collapse. The final answer to the questions raised here would be provided by the 3D numerical MHD codes
(see in \cite{Anderson2008,Giacomazzo2009} for the recent developments),
but this work provides hints and raises issues that need to be studied. Furthermore, as future studies,
we condiser to study magnetic effects on the gravitational waves emitted from the more complicated background
collapsing models such as inhomogeneous dust collapse and the stellar
collapse with perfect fluid, while it should be also important to take into account the background magnetic field
such as \cite{KSLS2009}.

\acknowledgments

We would like to thank K.D. Kokkotas and J.M. Mart\'{\i}n-Garc\'{\i}a for valuable comments.
This work was supported via the Transregio 7 ``Gravitational Wave Astronomy"
financed by the Deutsche Forschungsgemeinschaft DFG (German Research Foundation).

\appendix

\section{Concrete Expression for the Source Terms}   
\label{sec:appendix_1}

In this appendix, we show the concrete expression for the source terms in the perturbation equations,
which are not written in the main text. Those for Eqs. (\ref{perturbation-01}) -- (\ref{perturbation-03})
and (\ref{perturbation-05}) -- (\ref{perturbation-07}) are
\begin{align}
 S_\zeta =& 16\pi\left[T^3 + 2(n^A T_A)' - n^A n^B T_{AB} + 2(2\nu - W)n^A T_A\right]
     + 4(\mu - U)\dot{k} + 3\mu \dot{\zeta} + (2W-5\nu) \zeta' \nonumber \\
    &- 2(\mu - U) \dot{q} + (8\nu - 6W)q' + 2q''
     + 2\left[2(\mu - U)W - 2\mu\nu - \mu' + \dot{\nu} + 8\pi t^{AB}u_A n_B\right]\psi \nonumber \\
    &+ 2\left[2W' - 2\nu^2 + \frac{2}{r^2} - 16\pi Q + 8\pi (t^A_{\ A} + t^{AB} p_{AB})\right](\zeta + k)
     + 16\pi Q\zeta + \frac{(l-1)(l+2)}{r^2}\zeta  \nonumber \\
    &- 2\left[4W' + 4\nu W - 2W^2 - 2\nu^2 + \frac{l(l+1)+4}{2r^2}
      - 16\pi Q + 8\pi (t^A_{\ A} + t^{AB}p_{AB})\right]q, \\
 S_k     =& 8\pi \left[(-c_s^2 u^A u^B + n^A n^B)T_{AB} + 4W n^A T_A\right] \nonumber \\
    &+ (c_s^2 + 1)U \dot{\zeta} + \left[4U + c_s^2(\mu + 2U)\right]\dot{k}
      + (c_s^2 -1) W\zeta' - (\nu + 2c_s^2 W)k' - 2U\dot{q} + 2Wq' \nonumber \\
    &+ \Bigg[c_s^2\left\{8\pi (t^A_{\ A} - t^{AB}p_{AB}) + 2U(2\mu + U) + \frac{l(l+1)}{r^2}\right\}
     - 8\pi(t^A_{\ A}+t^{AB}p_{AB}) + 2\left(W^2 - \frac{1}{r^2}\right)\Bigg](\zeta + k) \nonumber \\
    &- \frac{(l-1)(l+2)}{2r^2}(c_s^2 + 1)\zeta 
     - \left[c_s^2\left\{8\pi(t^A_{\ A}-t^{AB}p_{AB}) + 2U(2\mu + U)\right\}
      - 8\pi(t^A_{\ A}+t^{AB}p_{AB}) + 6W^2 - \frac{l(l+1)+2}{r^2}\right]q \nonumber \\
    &+ 2\left[(c_s^2 + 1)U(\nu + W) + (c_s^2 -1)\mu W -8\pi (c_s^2 + 1) t^{AB}u_A n_B\right]\psi, \\
 S_\psi  =& 2\nu(\zeta+k) + 2\mu\psi + \zeta' - 2q' + 2(W-\nu)q - 16\pi n^A T_A,
\end{align}
and
\begin{align}
 C_{\gamma} =& - W\dot{\zeta} + U\zeta' + (2U - \mu)k' - 2Uq'
      + \left[\frac{l(l+1)+2}{2r^2} + U(2\mu + U) - W(2\nu + W) + 8\pi t^A_{\ A}\right]\psi, \\
 C_{\omega} =& U\dot{\zeta} + (\mu + 2U)\dot{k} + W\zeta' - 2Wk' \nonumber \\
    &+ \left[8\pi(t^A_{\ A} - t^{AB}p_{AB}) + 2U(2\mu + U) + \frac{l(l+1)}{r^2}\right](\zeta + k)
      - \frac{(l-1)(l+2)}{2r^2}\zeta \nonumber \\
    &- \left[8\pi(t^A_{\ A} - t^{AB} p_{AB}) + 2U(2\mu+U)\right]q
      + 2\left[\nu U + \mu W + UW - 8\pi t^{AB} u_A n_B\right]\psi, \\
 C_{\alpha} =& 2\mu (\zeta +k) + 2\nu \psi + \dot{\zeta} + 2\dot{k} - 2q (\mu + U),
\end{align}
where $q$ is given by Eq. (\ref{perturbation-04}), such as
\begin{equation}
 q = \frac{{b_1}^2-{b_3}^2}{\sqrt{5\pi}R^2}. \label{q}
\end{equation}
Additionally, the source terms in the perturbation equations for the metric perturbations
(\ref{OS-01}) -- (\ref{OS-03}) are
\begin{align}
 \bar{S}_\zeta &= 2\partial_\chi^2 q - \frac{6\cos\chi}{\sin\chi}\partial_\chi q + \frac{4\cos^2\chi - l(l+1) + 4}{\sin^2\chi}q
     + \frac{2}{\sqrt{5\pi}R^2\sin^2\chi}\left[\frac{2{b_2}^2}{\sin^2\chi} + {b_1}^2 + {b_3}^2 + 2\partial_\chi(b_1 b_2)
     - \frac{6\cos\chi}{\sin\chi}b_1b_2\right], \\
 \bar{S}_k &= \frac{2\partial_\eta R}{R}\partial_\eta q - \frac{2\cos\chi}{\sin\chi}\partial_\chi q
     + \frac{6\cos^2\chi - l(l+1) -2}{\sin^2\chi}q
     + \frac{1}{\sqrt{5\pi}R^2\sin^2\chi}\left[\frac{{b_2}^2}{\sin^2\chi} + {b_1}^2
     + {b_3}^2 - \frac{4\cos\chi}{\sin\chi}b_1b_2\right], \\
 \bar{S}_\psi &= 2\partial_\chi q - \frac{2\cos\chi}{\sin\chi}q + \frac{2 b_1b_2}{\sqrt{5\pi}R^2\sin^2\chi}.
\end{align}

\section{Matter Perturbations for interior region}   
\label{sec:appendix_2}

The matter perturbations are given by using the variables for metric and magnetic perturbations,
which are determined after calculation for evolution of metric perturbations.
With Eqs. (\ref{perturbation-05}) -- (\ref{perturbation-07}), those are
\begin{gather}
 \gamma = \frac{1}{8\pi\rho R^2}\left[\partial_\eta\partial_\chi k - \frac{\cos\chi}{\sin\chi}\partial_\eta\zeta
     + \frac{\partial_\eta R}{R}\partial_\chi\zeta + \frac{\partial_\eta R}{R}\partial_\chi k
     - \frac{2\partial_\eta R}{R}\partial_\chi q + \left\{\frac{l(l+1)}{2\sin^2\chi}
     + 3\left(\frac{\partial_\eta R}{R}\right)^2 + 1 - 4\pi\rho R^2\right\}\psi\right], \\
 \omega = \frac{1}{8\pi\rho R^2}\Bigg[-\partial_\chi^2 k + \frac{2\partial_\eta R}{R}\partial_\chi \psi
     + \frac{\partial_\eta R}{R}\partial_\eta \zeta + \frac{3\partial_\eta R}{R}\partial_\eta k
     + \frac{\cos\chi}{\sin\chi}\partial_\chi \zeta - \frac{2\cos\chi}{\sin\chi} \partial_\chi k   \nonumber \\
     + \left\{6\left(\frac{\partial_\eta R}{R}\right)^2 + \frac{l(l+1)}{\sin^2\chi} - 8\pi\rho R^2
       \right\} (\zeta + k)
     - \frac{(l-1)(l+2)}{2\sin^2\chi}\zeta - 6\left(\frac{\partial_\eta R}{R}\right)^2q   \nonumber \\
     + \frac{4\partial_\eta R}{R}\frac{\cos\chi}{\sin\chi}\psi
     + \frac{8\pi\kappa}{R^2\sin^2\chi}
     \left({b_1}^2 + {b_3}^2 - \frac{{b_2}^2}{\sin^2\chi}\right)\Bigg], \\
 \alpha = \frac{1}{16\pi\rho R}\left[\partial_\chi \psi + \frac{2\partial_\eta R}{R}(\zeta + k)
     + \partial_\eta\zeta + 2\partial_\eta k - \frac{4\partial_\eta R}{R}q\right].
\end{gather}


\end{document}